\RequirePackage{scrlfile}
\PreventPackageFromLoading{natbib}

\documentclass[final,3p,times,numbers,compress,sort]{elsarticle}

\makeatletter
\def\ps@pprintTitle{%
  \let\@oddhead\@empty
  \let\@evenhead\@empty
  \def\@oddfoot{}
  \let\@evenfoot\@oddfoot
}
\makeatother

\makeatletter
\let\c@author\undefined
\makeatother

\usepackage{amssymb}
\usepackage{amsmath}
\usepackage{multirow}

\usepackage{lineno}
\usepackage{caption}
\usepackage{subcaption}
\usepackage{pdfpages}

\def\equationautorefname~#1\null{Eq.~(#1)\null}

\usepackage[backend=biber, style=authoryear, maxcitenames=1,mincitenames=1,uniquelist=false,maxbibnames=99,giveninits=true,isbn=false,url=false]{biblatex}

\DeclareNameAlias{sortname}{family-given}

\DeclareDelimFormat{multinamedelim}{\addcomma\space}
\DeclareDelimFormat{finalnamedelim}{\addcomma\space}

\DeclareDelimFormat[bib]{nameyeardelim}{\addcomma\space}   
\DeclareDelimFormat[bib]{nametitledelim}{\addperiod\space} 

\renewbibmacro*{date+extradate}{%
  \iffieldundef{labelyear}
    {}
    {\printtext{\printlabeldateextra}}}
    
\DeclareFieldFormat[article,book,incollection,inproceedings,thesis]{title}{#1}

\DeclareFieldFormat{titlecase}{%
  \ifcurrentfield{title}
    {\MakeSentenceCase*{#1}}
    {#1}}

\DeclareFieldFormat{journaltitle}{#1}
\DeclareFieldFormat{shortjournal}{#1}

\renewbibmacro*{in:}{%
  \ifentrytype{article}{}{\printtext{\bibstring{in}\intitlepunct}}%
}

\usepackage{xcolor} 
\usepackage{hyperref}

\hypersetup{
    colorlinks=true,       
    citecolor=teal,        
    linkcolor=red,         
    urlcolor=teal          
}

\DeclareFieldFormat{citehyperref}{%
  \DeclareFieldAlias{bibhyperref}{noformat}%
  \bibhyperref{#1}}

\savebibmacro{cite}
\savebibmacro{textcite}

\renewbibmacro*{cite}{%
  \printtext[citehyperref]{%
    \restorebibmacro{cite}%
    \usebibmacro{cite}}}

\renewbibmacro*{textcite}{%
  \ifboolexpr{
    test {\ifnumcomp{\value{citecount}}{=}{1}}
    and
    not test {\iffastcite}
  }
    {\printtext[citehyperref]{%
       \restorebibmacro{textcite}%
       \usebibmacro{textcite}}}
    {\restorebibmacro{textcite}%
     \usebibmacro{textcite}}}

\DeclareDelimFormat[parencite,cite]{nameyeardelim}{\addcomma\space}

\journal{}

\begin{document}

\begin{frontmatter}

\title{Fracture of Lattice Materials from Low to High Relative Density}

\author[UPenn]{Adam P. Taylor} 
\author[UPenn]{Sage Fulco}
\author[UPenn]{Kevin T. Turner\corref{cor1}}
\ead{kturner@engineering.upenn.edu}
\cortext[cor1]{Corresponding author.}

\address[UPenn]{Dept. of Mechanical Engineering and Applied Mechanics, University of Pennsylvania, Philadelphia, Pennsylvania 19104, United States}

\begin{abstract}
Lattice materials have highly tunable mechanical properties that are controlled by their internal geometry and base material behavior. While fracture models for low-density lattices are well-established, the fracture of lattices at relative densities above 30$\%$ remains largely unexplored. Here, we present an analytical model that combines homogenization methods and blunt-crack fracture theory to predict the fracture toughness of triangular and hexagonal lattices over a wide range of relative densities. Finite element modeling corroborates the analytical framework and demonstrates a shift in failure mechanisms between low and high relative density regimes. At high relative densities, failure is dominated by nodal stress concentrations, which are strongly influenced by the local crack-tip geometry and the macroscopic crack path. Significant stress redistribution is observed as the relative orientation and lattice fillet radius are varied, resulting in increased fracture toughness relative to baseline geometries. Furthermore, quasi-brittle poly(methyl methacrylate) (PMMA) lattices achieve about a 1.6$\times$ enhancement in fracture toughness relative to perfectly brittle lattices as failure is delayed by localized plasticity. Notably, some quasi-brittle lattices are found to exceed the fracture toughness of their base material -- a feature only possible when the cell size is large compared to the base material's plastic radius. Experiments on lattice fracture specimens made from laser-cut PMMA validate the finite element results, but highlight the stochastic nature of failure in brittle lattice materials, especially at higher relative densities.
\end{abstract}




\end{frontmatter}



\section{Introduction}
\label{intro}
The development of materials with tailored internal geometries has expanded the traditional property space of structural solids and enabled fine control over their mechanical response. Lattice materials have garnered significant interest over the past several decades, in part due to their impressive mechanical properties at low densities. In general, their properties depend on the properties of the chosen base material and on their internal architecture, two independent features whose effect on stiffness \parencite{zheng2014ultralight,deshpande2001effective,mueller2018architected}, strength \parencite{bauer2016approaching,meza2014strong,deshpande2001effective}, and fracture toughness \parencite{shaikeea2022toughness,fulco2025disorder,fleck2007damage} has been investigated by many. This has led to the development of materials that can be designed to have a wide range of mechanical properties. Advances in additive manufacturing techniques have facilitated the fabrication of structures with fine-scale architecture, unlocking their use in a broader range of mechanical systems \parencite{schaedler2011ultralight,chen2020size,harris2020metallic,geng2019mechanical}. However, the design space of lattices has not been fully explored. Most works to date investigating lattice fracture assume linear-elastic, brittle failure and are limited to relative densities below 30$\%$ (30$\%$ material and 70$\%$ void-space). More work is necessary to fully understand their range of fracture properties. Specifically, three avenues for making lattice materials tougher are frequently overlooked: 1) increasing relative density past 30$\%$, 2) utilizing material plasticity, and 3) controlling the crack path via changes in the relative orientation between the lattice geometry and the loading direction.

The properties of a lattice depend strongly on the specific unit cell geometry. A lattice can be classified as stretch-dominated or bending-dominated depending on the connectivity, leading to very different mechanical responses \parencite{gibson1997cellular,ashby1983mechanical,maiti1984fracture,gibson1981elastic}. The fracture properties of a 2D honeycomb lattice were first investigated by Gibson, Ashby, and others in the early 1980s, who showed that fracture toughness depended strongly on relative density, $\bar{\rho}$, a dimensionless quantity relating the lattice density to the base material density \parencite{gibson1997cellular,gibson1982mechanics2D}. More recently, the fracture properties of lattices with many different cell geometries have been investigated, including lattices with both stretch- and bending-dominated behaviors \parencite{tankasala20152013,fleck2010micro}, geometrical disorder \parencite{schmidt2001ductile,fulco2025disorder}, negative Poisson's ratio \parencite{lakes1987foam,gibson1982mechanics3D}, and structural hierarchy \parencite{fleck2010micro,leraillez2025fracture}. However, ordered triangular and hexagonal lattices are commonly used as model geometries as they represent the two main classes of lattices. For relative densities below 30$\%$, it has been shown that lattice fracture toughness scales with $\bar{\rho}$ according to: $K_\textnormal{Ic}\propto\bar{\rho}^d$, where $d$ depends on the lattice cell geometry (e.g., triangular, hexagonal, kagome) \parencite{ashby1983mechanical,maiti1984fracture,gibson1981elastic}. This indicates that the toughest lattices are at the highest densities, but the assumptions made within these lattice fracture models become invalid in this regime, leaving it unclear whether the model predictions remain accurate \parencite{gibson1997cellular,huang1991fracture}. A recent work looked at the fracture behavior of lattices at low slenderness ratios (a parameter that is inversely related to relative density), but the analysis focuses on the effects of material stochasticity, and the numerical investigations use a beam model to represent the lattice, which does not fully capture the local stress field at higher densities \parencite{chouzouris2026geometry}. Another recent study investigated how relative density affects the strength and modulus of lattices, but does not consider the effects on fracture at these high densities \parencite{emami2026mechanical}. Classical lattice fracture models assume that when the strut aspect ratio is small (when $\bar\rho<0.2$, generally), then each strut in the lattice can be modeled as a beam to facilitate simple calculation of their internal stresses. As this ratio increases, however, the stress state becomes more complex, and the internal stress distribution of each strut is not well described by the beam model, which captures only axial and bending stresses. At very high relative densities, the lattice more closely resembles an architected material containing a pattern of holes, with crack tip stresses more akin to those of a homogeneous material with a blunt crack \parencite{torabi2014closed,bijaya2023multiscale}. This change in geometry suggests a transition in fracture behavior and fracture toughness between lattice materials with low and high relative densities.

The fracture toughness of a lattice is also limited by the properties of its base material -- selecting a tougher base material will generally produce a tougher lattice \parencite{fleck2010micro}. Many studies on lattice fracture assume elastic-brittle failure, which significantly limits their fracture properties, since a lattice with an elastic-brittle base material will fail as soon as any material point reaches the fracture strength. However, many common engineering materials exhibit some amount of plasticity, a mechanism that can improve the fracture toughness of lattices by softening local stress concentrations near the crack tip and dissipating additional energy during crack propagation. Using plasticity to improve the fracture toughness of 2D and 3D lattices has been shown in the literature through numerical and experimental studies \parencite{olurin2000deformation,schmidt2001ductile,ajdari2008effect,tankasala20152013,manno2019engineering,tankasala2020crack,hsieh2020versatile,gu2019fracture,o2017fracture,maurizi2022fracture,choukir2023interplay,fulco2024enhancing,fulco2022decoupling}. Schmidt and Fleck \parencite{schmidt2001ductile} first solved for the toughness and plastic zone shape in a regular hexagonal lattice and found that increasing the fracture strength and decreasing the hardening modulus of the parent material lead to a tougher lattice. Later, models for the stress intensity factor of 2D elastic-plastic lattices with different cell shapes (triangle, hexagon, diamond, and kagome) were derived as a function of the strain hardening exponent, $n$, and strain to failure, $\varepsilon_f$, of the parent material \parencite{tankasala20152013}. More recently, 3D elastic-plastic lattices have gained significant attention as highly tough, lightweight materials that can be fabricated at very small scales \parencite{maurizi2022fracture}. However, introducing material plasticity in lattices with higher relative densities (above 30\%) has not been systematically explored, and its influence on the fracture process is unclear.

Triangular and hexagonal lattices have relatively isotropic fracture properties -- their fracture toughness does not depend on the relative angle between the geometry and the loading direction. Most lattice geometries exhibit less than 15\% difference in fracture toughness across all relative orientations as shown through a number of analytical, numerical, and experimental studies \parencite{chen1998fracture,lipperman2007fracture,gu2018experimental,bijaya2023multiscale}. Thus, little to no improvement in fracture toughness can be expected if the orientation is changed. However, these studies generally consider lattices with $\bar{\rho}< 30\%$. At higher relative densities, the stress distribution in each strut will change and may become more sensitive to the local geometry \parencite{taylor2025hinged}. In high relative density porous media ($\bar{\rho}>0.7$, generally), a dependence on orientation \textit{has} been observed \parencite{bijaya2023multiscale}. There is likely a relative density at which relative orientation begins to significantly influence failure near the crack tip, leading to anisotropic fracture toughness.

Building on the well-established fracture models for lattice materials, the objective of this work is to investigate the role of relative density, plasticity, and relative orientation on the fracture of lattice materials. Taylor et al. showed experimentally that as $\bar{\rho}$ increases past 0.3, generally, fracture toughness increases monotonically up to $\bar{\rho}=0.8$, but deviates significantly from classical lattice fracture models \parencite{taylor2025hinged}. The origins of this behavior were not fully explored in that work, highlighting the importance of investigating higher-density lattices. This work presents an analytical model to predict lattice toughness across a range of relative densities and introduces a finite element (FE) framework to calculate lattice fracture toughness. The FE framework includes two failure criteria to predict the failure location and peak load (perfectly-brittle and quasi-brittle criteria), as well as a method to identify the fracture path through each lattice cell. An energy-based fracture analysis is then performed to extract lattice fracture toughness from the FE results. FE results for a wide range of relative densities, orientations, and multiple material behaviors are compared to the analytical predictions and validated through experiments on laser-cut PMMA lattices.

\section{Analytical Framework and Modeling}
\label{theory}

The fracture behavior of high relative density lattices was observed experimentally in our recent work, where fracture toughness is shown to increase monotonically as relative density is increased up to $\bar{\rho}\approx0.8$, but classical lattice fracture models were not sufficient for predicting lattice toughness (and tend to over predict) \parencite{taylor2025hinged}. Additionally, failure occasionally occurred through nodes rather than struts in some triangular geometries, which is not captured by traditional lattice fracture models and has not been fully studied in the literature. To model the fracture process of high-$\bar{\rho}$ lattice materials, we first present a framework that is incorporated within FE simulations to 1) determine the failure point (one failure criterion which assumes elastic-brittle failure and one which assumes quasi-brittle failure), 2) predict the path of a crack through a unit cell, and 3) calculate the fracture toughness of a lattice with a deflected crack path. This computational model is used to evaluate the effect of relative orientation, density, and material plasticity on the fracture process. Additionally, we propose an analytical fracture model that homogenizes the lattice geometry and employs blunt-crack fracture theory to estimate the local stresses that develop at the lattice crack tip and predict lattice toughness.

\subsection{Model Geometry}

\begin{figure}[h]
\centering
\includegraphics[width=1\textwidth,trim={0.2cm 0.2cm 0.2cm 0.2cm},clip]{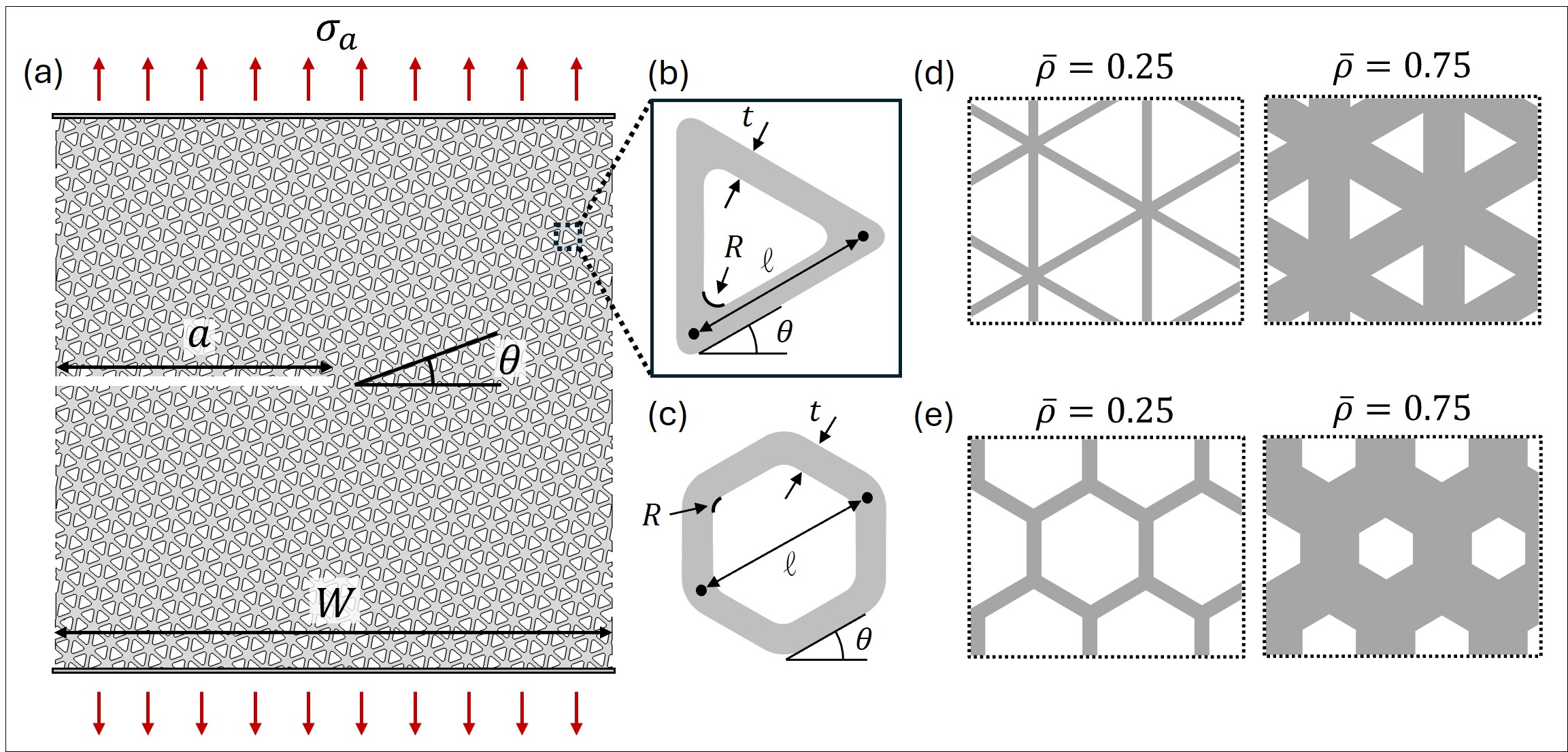}
\caption{Overall specimen geometry. (a) Single-edge notch tension fracture specimen containing the lattice structure with a precrack, $a$, and relative orientation, $\theta$. The unit cell geometry of a (b) triangular and (c) hexagonal lattice. Example geometries of (d) triangular and (e) hexagonal lattices at a low (25$\%$) and high (75$\%$) relative density.}\label{Fig1}
\end{figure}

We classify a lattice as low-$\bar{\rho}$ when $\bar{\rho}<0.3$ and high-$\bar{\rho}$ otherwise. Triangular and hexagonal unit cells with associated cell size, $\ell$, strut thickness, $t$, and relative angle (between the lattice material orientation and the loading direction), $\theta$, are used to construct lattice structures with varying density and orientation as shown in \autoref{Fig1}. Cell size is held constant throughout the study while $t$ is varied to induce changes in relative density, as shown in \autoref{Fig1}(d-e). Fillets are added to the structure to prevent unrealistic stress concentrations at the nodes. We note here that nominal relative densities are calculated for each lattice in this work -- the material contribution from the fillets towards relative density is ignored. The overall specimen is a square with a precrack at the midplane and has a crack length $a=W/2$, where $W$ is the specimen width. The crack tip is located at the center of the cell that is closest to the center point of the specimen. The overall number of unit cells of the lattice is fixed at 25$\times$25 in order to generate a K-field at the crack tip that is small compared to the size of the specimen. For specimens larger than the size used, fracture toughness measurements become independent of specimen size, so the fracture toughness obtained from this 25$\times$25 geometry is taken as the effective fracture toughness of the lattice. These lattice specimen dimensions also follow the general rule for lattice fracture analyses, $a/\ell>7$, which ensures toughness-controlled failure \parencite{hsieh2020versatile,gibson1997cellular,huang1991fracture}. The far right boundary will begin to affect the K-field at large crack lengths, so we only use crack lengths of $a < 0.8W$. Details of the range of dimensions for each geometric parameter are described in \autoref{femodel}.

\subsection{Modeling Lattice Failure Initiation}\label{latticeFailureMethods}
Modeling the failure of lattice materials has traditionally been performed by assuming each strut in the lattice can be approximated as a beam under some combination of axial and bending stresses \parencite{gibson1997cellular,huang1991fracture}. These models calculate the maximum stress in a strut near the crack tip and compare it to the tensile strength of the base material to assess failure and find lattice fracture toughness. Lattice fracture toughness, $K_{\textnormal{Ic}}$, scales with $\bar{\rho}$ and $\sqrt{\ell}$ at low-$\bar\rho$:
\begin{equation}
    {K_{\textnormal{Ic}}} = D\:\bar{\rho}\,^{d}\sigma_{\textnormal{TS}}\sqrt{\ell}
\label{eq:KIclattice}
\end{equation}
where $D$ and $d$ are scaling coefficients that depend on the cell shape ($D=0.5$, $d=1$ for triangles, and $D=0.9$, $d=2$ for hexagons), and $\sigma_{\textnormal{TS}}$ is the tensile strength of the parent material \parencite{gibson1997cellular,fleck2010micro}. This fracture model belongs to a family of models, commonly referred to as the Gibson-Ashby (GA) models, that predict the mechanical properties of lattice materials. The GA models have been validated both computationally and experimentally and are frequently used to estimate the properties of lattices at relative densities below 30$\%$ \parencite{gibson1997cellular,fleck2007damage}.

However, when $\bar{\rho}>0.3$, generally, and for a fixed, small fillet radius, the assumption that the stresses within each strut can be modeled using beam theory break down \parencite{taylor2025hinged}. Stress concentrations at each node can control failure in these high-$\bar{\rho}$ lattices, yet no models exist to predict their magnitude. Moreover, the mechanical properties of the parent material (i.e., having perfectly elastic-brittle or quasi-brittle behavior) and the local crack tip geometry may also significantly affect how these stress concentrations contribute to failure.

\subsubsection{Elastic-Brittle Failure}

The first and simplest failure criterion used here is the elastic-brittle (EB) failure criterion -- failure occurs when the maximum principal stress, $\sigma_1$, anywhere in the lattice equals the yield strength of the parent material, $\sigma_y$, assuming $\sigma_y\approx\sigma_{\textnormal{TS}}$. Most lattices do not have a sharp crack tip, so the stresses are not singular, and the calculation of lattice fracture toughness with elastic-brittle properties is tractable. For low-$\bar{\rho}$ lattices, beam theory-based models accurately predict the maximum stress within a strut in the lattice. Small stress concentrations near nodes may still develop in stretch-dominated, low-$\bar{\rho}$ lattices that these models do not capture, leading to slightly higher predicted values of $K_{\textnormal{Ic}}$ than in reality, but this effect is considered negligible \parencite{quintana2009fracture}. As shown below, as $\bar{\rho}$ increases beyond 30$\%$, the stress distribution in the strut will change, with stress concentrating at the node, but failure can still be described by the elastic-brittle criterion.

\subsubsection{Circular Notch Fracture of a Quasi-Brittle Material}\label{bluntCrackFrac}

\begin{figure}[h]
\centering
\includegraphics[width=1\textwidth-2cm,trim={0.2cm 0.2cm 0.2cm 0.2cm},clip]{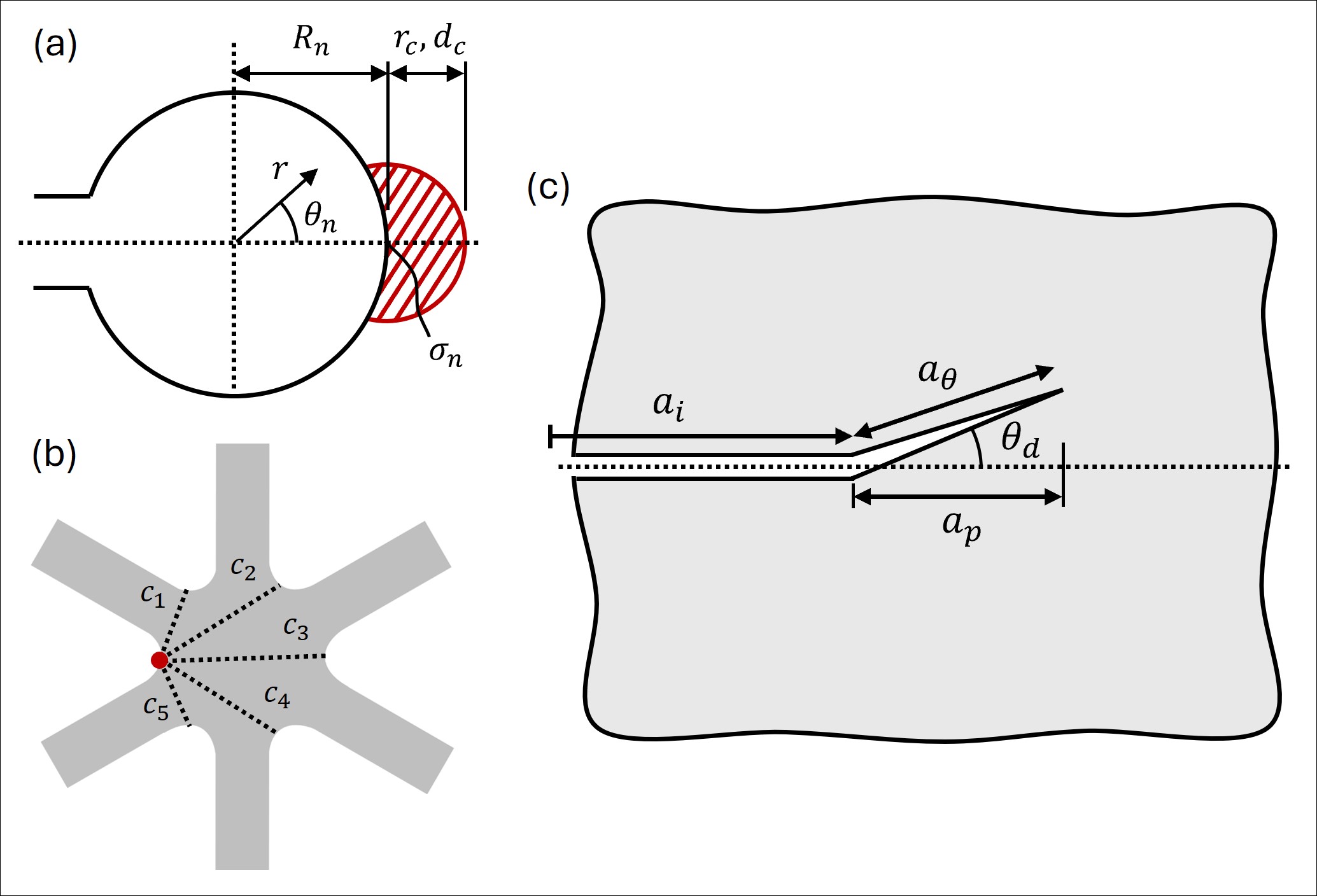}
\caption{Schematics for computational modeling criteria and methods. (a) A circular notch crack tip with representative plastic zone geometry for a quasi-brittle material. (b) Potential crack paths in a triangular lattice cell used by the MERR criterion. (c) Geometry of a deflected crack used in the compliance method for calculating lattice toughness.}\label{Fig2}
\end{figure}

Most materials are not perfectly elastic-brittle and exhibit some amount of plasticity. A material with \textit{nearly} elastic-brittle behavior is said to be \textit{quasi}-brittle (QB). Some examples of QB materials include brittle polymers such as PMMA and high-strength, low-ductility metallic materials \parencite{torabi2014closed}. A quasi-brittle \textit{lattice} will fail abruptly at the location of maximum stress, but some amount of localized plasticity that delays failure is expected. To model QB failure in a lattice, we use an approach previously developed to model the quasi-brittle fracture of circular notch specimens \parencite{creager1967elastic,kullmer2006influence,torabi2014closed}. This model is well-suited for modeling fracture in lattices because the crack tip has a finite radius. In the model, the elastic stress distribution ahead of a circular notch fracture specimen follows the typical $1/\sqrt{r}$ scaling (consistent with classical fracture mechanics), but the stress distribution is shifted to account for the blunt crack tip. The tangential stress is given along the line that bisects the circular notch specimen:
\begin{equation}
    \sigma_{t}(r,\theta_n=0^\circ)=\frac{K_{\textnormal{I}}^{\textnormal{n}}}{2\sqrt{2\pi r}}\left[ 2+1.25\left(\frac{R_n}{r}\right)+1.5\left(\frac{R_n}{r}\right)^2+1.25\left(\frac{R_n}{r}\right)^3\right]
\label{eq:elasNotchStress}
\end{equation}
where ${K_{\textnormal{I}}^{\textnormal{n}}}$ is the mode-I notch stress intensity factor, $R_n$ is the notch radius, and $r$ and $\theta_n$ are the radius and angle from the center of the circular notch, respectively, as shown in \autoref{Fig2}(a). The tangential stress at the notch edge, $\sigma_n$, is at $r=R_n$ and $\theta_n=0^\circ$, and is equal to: 
\begin{equation}
    \sigma_n=3\frac{K_{\textnormal{I}}^{\textnormal{n}}}{\sqrt{2\pi R_n}}.
\label{eq:sigTanNotch}
\end{equation}
The notch fracture toughness, $K_{\textnormal{Ic}}^{\textnormal{n}},$ of a perfectly elastic-brittle material is determined from \autoref{eq:sigTanNotch}, assuming $\sigma_n=\sigma_{\textnormal{TS}}$, leading to:
\begin{equation}
    K_{\textnormal{Ic}}^{\textnormal{n}}=\frac{1}{3}\sigma_{\textnormal{TS}}\sqrt{2\pi R_n} .
\label{eq:sigyyNotch}
\end{equation}

To approximate failure of a \textit{quasi-brittle} material, a point stress (PS) or mean stress (MS) criterion can be applied \parencite{torabi2014closed}. The criteria account for the limited plasticity that develops ahead of the notch by introducing a critical distance parameter. The critical distances $r_c$ and $d_c$ for the PS and MS criteria, respectively, are a function of $\sigma_{\textnormal{TS}}$ and the fracture toughness of the base material, $K_{\textnormal{Ic,b}}$, and define the point or region of interest for evaluating the elastic stress distribution (\autoref{eq:elasNotchStress}). The PS criterion uses the stress located a distance, $r_c$, from the edge (\autoref{Fig2}(a)), where:
\begin{equation}
    r_c=\frac{1}{2\pi}\left(\frac{K_{\textnormal{Ic,b}}}{\sigma_{\textnormal{TS}}}\right)^2
\label{eq:rc}
\end{equation}
while the MS criterion uses the average stress across a distance, $d_c$, from the edge, where:
\begin{equation}
    d_c=\frac{2}{\pi}\left(\frac{K_{\textnormal{Ic,b}}}{\sigma_{\textnormal{TS}}}\right)^2.
\label{eq:dc}
\end{equation}
Quasi-brittle failure is predicted to occur when the point stress at $r_c$ or the average stress over $d_c$ reaches the tensile strength of the parent material, $\sigma_{\textnormal{TS}}$. In general, the MS criterion tends to predict a slightly higher fracture toughness than the PS criterion, but both have been used with good accuracy and validated experimentally on PMMA \parencite{torabi2014closed}.

If the stress distribution ahead of the nodal stress concentrations in a lattice is similar to the stress distribution of a circular notch specimen, then failure can be described by the PS or MS criterion. The smallest length scale of the lattice (the strut thickness) should be large compared to the critical distance ($d_c$ or $r_c$) to ensure the stress distribution is unaffected by the local geometry. The MS criterion is used whenever possible, as it averages over the elastic stress distribution, making it less sensitive to numerical noise. The PS criterion is used when the MS critical distance becomes comparable to the lattice strut size. 

\subsection{Modeling Lattice Fracture}\label{crackGrowthMethod}

After a crack initiates in the lattice, it will travel along some path through the cell. For low-$\bar{\rho}$ lattices, a crack will most likely grow through the strut that carries the highest stress as is assumed by the Gibson-Ashby models. As $\bar{\rho}$ increases and as the orientation of the unit cell changes, however, the crack path through the cell becomes less clear. Nodal failure occurs when a crack grows through the node and disconnects multiple struts, a behavior that is only possible in lattices with connectivity greater than 3. This has recently been experimentally observed in high-$\bar{\rho}$ PMMA lattices and enables the generation of new crack paths \parencite{taylor2025hinged}. An energy-based method is proposed to predict the direction a crack will grow through the cell of a lattice of any relative density. Then, the effective toughness of the lattice, $G_c$, can be calculated using energy-based methods and converted to an effective fracture toughness, $K_\textnormal{Ic}$, using the Irwin relation: $G_c =K_{\textnormal{Ic}}^2/E^*$, where $E^*$ is equal to specimen modulus, $E$, for plane stress and $E/(1-\nu^2)$ for plane strain \parencite{irwin1957analysis}. We assume plane strain for this analysis as the relevant length scale to compare to the specimen thickness, $b$, is the strut thickness. For convenience, lattice effective toughness and effective fracture toughness will henceforth be referred to as toughness and fracture toughness, respectively.

\subsubsection{The Maximum Energy Release Rate (MERR) Crack Path}

For a homogeneous material with a sharp crack tip, the direction a crack grows depends upon which path minimizes the total energy of the system. This basic idea is used in phase-field modeling and allows the explicit calculation of a moving crack through many material systems \parencite{francfort1998revisiting}. While phase-field methods can be accurate and robust, they are often computationally expensive and difficult to implement effectively. Another way to approach the problem is to choose the crack path which maximizes the strain energy release rate of the system \parencite{hussain1974strain}. Starting from Griffith's theory of fracture and assuming a displacement-control boundary condition, a crack will grow when the following inequality is satisfied:
\begin{equation}
    -\frac{dU}{dA}=G\geq G_c
\label{eq:G-LEFM}
\end{equation}
where $U$ is the elastic strain energy, $A$ is the crack area, $G$ is the strain energy release rate, and $G_c$ is the critical value of the strain energy release rate at failure, a material property. The crack path that minimizes the total energy of the system will also be the first to satisfy the above equations such that $G=G_c$. This path is the most energetically favorable for the crack to grow along. In a lattice, the plausible paths a crack will take through a cell are limited and depend on the cell geometry. Assuming the crack initiates at the point of highest stress (the red point in \autoref{Fig2}(b)), and grows instantaneously through the cell such that the stress state around the crack tip isn't able to update as it travels, the crack will grow along one of the five straight paths, $c_1$ through $c_5$, as shown in \autoref{Fig2}(b). The actual crack path through a real lattice unit cell may be nonlinear, but it will be nearly energetically identical to one of the five straight cracks shown. Note that paths $c_1$ and $c_5$ correspond to strut failures while paths $c_2$ through $c_4$ correspond to nodal failures. If the change in elastic strain energy, $\Delta U$, and corresponding increase in crack area, $\Delta A$, are known for each path, then the path with the maximum $-\Delta U/\Delta A$ is the one the crack is most likely to follow. 

\subsubsection{Lattice Toughness Calculations}\label{angComMethod}

A linear-elastic fracture mechanics (LEFM) framework is used to calculate lattice toughness. In LEFM, the strain energy release rate can be written as a function of system compliance, $C$, such that:
\begin{equation}
    G=\frac{P^2}{2b}\frac{dC}{da}
\label{eq:Gcompliance}
\end{equation}
where $P$ is the applied load, $b$ is the specimen thickness, and $a$ is the crack length. The $C$-$a$ curve can be calculated analytically or through computational methods for the specific specimen geometry, so that only either $C$ or $a$ needs to be measured during testing to evaluate $dC/da$. The $C$-$a$ equation will depend on the in-plane geometry, specimen thickness, $b$, and specimen modulus, $E$. In the case of a lattice, an effective modulus, $E_f$, may be calculated computationally. Experimentally, system compliance is commonly measured during unloading or reloading at different crack lengths along with peak loads, $P_c$, to calculate toughness -- a procedure known as the compliance method. However, the compliance method typically assumes that the crack will propagate straight along the specimen's midplane. For cases where the crack deflects, as is often the case in lattice materials, a modification is introduced to allow for the compliance method to be applied to angled cracks. 

The $C$ vs. $a$ curve for a straight crack is expected to follow a fifth-order polynomial as seen in other single-edge notch tension (SENT) specimen geometries \parencite{tada2000stress}:
\begin{equation}
C E_f=
        D_1\left(\frac{a_{p,t}}{W}\right)^5 
        +D_2\left(\frac{a_{p,t}}{W}\right)^4  
        +D_3\left(\frac{a_{p,t}}{W}\right)^3 
        +D_4\left(\frac{a_{p,t}}{W}\right)^2  
        +D_5\left(\frac{a_{p,t}}{W}\right)
        +D_6
\label{eq:CvsAcurve}
\end{equation}
where $D_1$ through $D_6$ are fitting coefficients determined computationally, $W$ is the specimen width, and $a_{p,t}$ is the length of a straight crack growing along the midplane of the specimen. To construct the $C$ vs. $a$ curve of a specimen with a \textit{deflected} crack, we assume that the compliance of a specimen with a deflected crack (like shown in \autoref{Fig2}(c)) that has precrack length, $a_i$, deflection angle, $\theta_d$, and deflected crack length, $a_\theta$, is equal to that of a specimen with a straight crack of total length $a_{p,t}=a_i+a_p$, where $a_p$ is the projection of $a_\theta$ onto the specimen centerline. The amount of uncracked material remains constant in both cases (for deflected or projected cracks) and governs the system's stiffness. This assumption simplifies the FE study necessary to develop an equation for specimen compliance as a function of the crack geometry and is validated in a separate FE study. 

The toughness of a specimen with a deflected crack is found by considering the change in specimen compliance with respect to the change in \textit{real} crack length, or $dC/da_{\theta,t}$. The total length of the deflected crack, $a_{\theta,t}$, can be found knowing the final crack geometry ($a_{p,t}$, $a_i$ and $\theta_d$), using:
\begin{equation}
    a_{p,t}= a_i+(a_{\theta,t}-a_i)\cos(\theta_d).
\label{eq:aptEq}
\end{equation}
Substituting for $a_{p,t}$ in \autoref{eq:CvsAcurve} using \autoref{eq:aptEq}, taking the derivative with respect to $a_{\theta,t}$, and substituting the resulting $dC/da_{\theta,t}$ equation into \autoref{eq:Gcompliance} produces an equation for the strain energy release rate as a function of $a_{\theta,t}$, $P$, and the specimen geometry and modulus. Material toughness, $G_c$, is calculated by measuring peak loads, $P_c$, at rupture and using measurements of system compliance to calculate $a_{\theta,t}$ from \autoref{eq:CvsAcurve} and \autoref{eq:aptEq}, knowing $a_i$ and $\theta_d$ from the final macroscopic crack paths. This method can be applied to all rupture events during lattice fracture to obtain multiple toughness measurements. The results can also be averaged to obtain steady-state values and converted to fracture toughness, $K_\textnormal{Ic}$, using the Irwin relation to enable comparisons across all lattice geometries.

\subsection{Homogenized Lattice Fracture Theory}\label{lattHomogTheory}

\begin{figure}[h]
\centering
\includegraphics[width=1\textwidth-6cm,trim={1.5cm 2.5cm 1cm 0.2cm},clip]{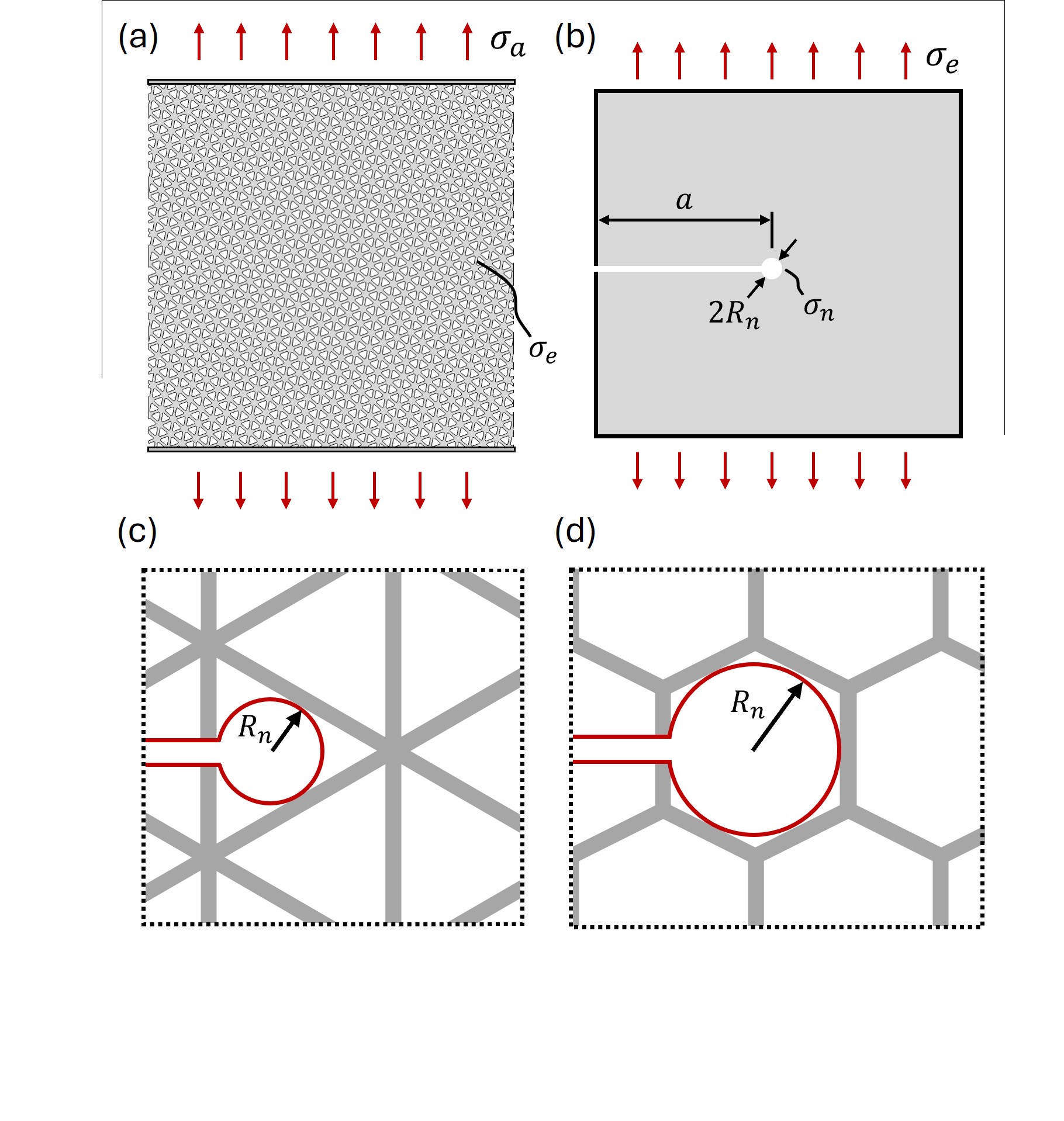}
\caption{Homogenized lattice model geometry and stresses. (a) The uncracked lattice geometry is loaded with an applied stress, $\sigma_a$, and we define an equivalent lattice stress, $\sigma_e$, that is a function of the lattice geometry and $\sigma_a$. (b) The homogenized lattice fracture geometry with a notched precrack. The effective circular notch radii for a (c) triangular and (d) hexagonal lattice.}\label{Fig3}
\end{figure}

The previous sections introduced a framework to calculate the fracture toughness of lattices at any relative density, but require the implementation of computational methods. To predict fracture toughness analytically, we treat the lattice as a homogeneous solid with effective properties and again use blunt crack fracture theory as presented previously in \autoref{bluntCrackFrac}, to estimate the high stresses at the crack tip. A homogenization is first applied to an \textit{uncracked} lattice, like the one shown in \autoref{Fig3}(a), to approximate the average stress distribution across the specimen when loaded by an applied stress, $\sigma_a$. For the stretch-dominated triangular lattice, we assume axial stresses dominate the stress distribution in each strut, even at high densities, so an equivalent lattice stress, $\sigma_{e,\textnormal{tri}}$, can be calculated knowing the cross-sectional area of the lattice:
\begin{equation}
    \sigma_{e,\textnormal{tri}}=\frac{\sigma_a A_o}{A_\textnormal{latt}}=\frac{\sigma_a}{\bar{\rho}}
\label{eq:sigEqTri}
\end{equation}
where $A_o$ is the nominal specimen cross-sectional area, and $A_\textnormal{latt}$ is the real lattice cross-sectional area.

In the bending-dominated hexagonal lattice, the internal stress state is more complex and depends on the angle between the strut axis and the loading direction, as well as on relative density. However, in the $\theta=30^\circ$ case where some struts are aligned with the loading direction, as seen in \autoref{Fig1}(e), the stress distribution can be approximated from the axial stress that forms in these vertical struts. The equivalent stress, $\sigma_{e,\textnormal{hex}}$, again depends on the ratio between the whole specimen cross-sectional area and the real lattice area along the cross-section that bisects the vertical struts:
\begin{equation}
    \sigma_{e,\textnormal{hex}}=\frac{\sigma_a A_o}{A_\textnormal{latt}}=\frac{\sigma_a N (\frac{\sqrt{3}\ell}{2})}{Nt}=\frac{\sqrt{3}}{2}\frac{\sigma_a \ell}{t}
\label{eq:sigEqHex}
\end{equation}
where $N$ is the number of unit cells along the specimen width. This model is expected to generalize across all orientations as the hexagonal lattice is considered isotropic \parencite{gibson1997cellular}.

To produce the same internal stress state in a homogeneous material, the equivalent stresses derived in \autoref{eq:sigEqTri} and \autoref{eq:sigEqHex} are applied at the far-field as seen in \autoref{Fig3}(b). Then, to capture the local stresses that develop near the crack tip in the lattice, a blunt crack with length $a$ and tip radius, $R_n$, is added to the homogeneous specimen. Circular notch fracture theory, as presented in \autoref{bluntCrackFrac}, can then be applied to predict the fracture toughness of the lattice. The notch radius will depend on the cell geometry of the lattice as shown in \autoref{Fig3}(c-d). A circle is inscribed in the unit cell of both a triangle and hexagon to estimate their effective notch radii, $R_n$. Both notch radii are functions of the unit cell size, $\ell$, and strut thickness, $t$:
\begin{equation}
    R_{n,\textnormal{tri}}=\frac{\ell}{2\sqrt{3}}-\frac{t}{2}
\label{eq:effRadTri}
\end{equation}

\begin{equation}
    R_{n,\textnormal{hex}}=\frac{3\ell}{4\sqrt{3}}-\frac{t}{2}
\label{eq:effRadHex}
\end{equation}
and converge to zero at $\bar{\rho}=1$, producing an infinitely sharp crack tip.

\begin{figure}[h]
\centering
\includegraphics[width=1\textwidth-1cm,trim={0.3cm 0.2cm 0.3cm 0.3cm},clip]{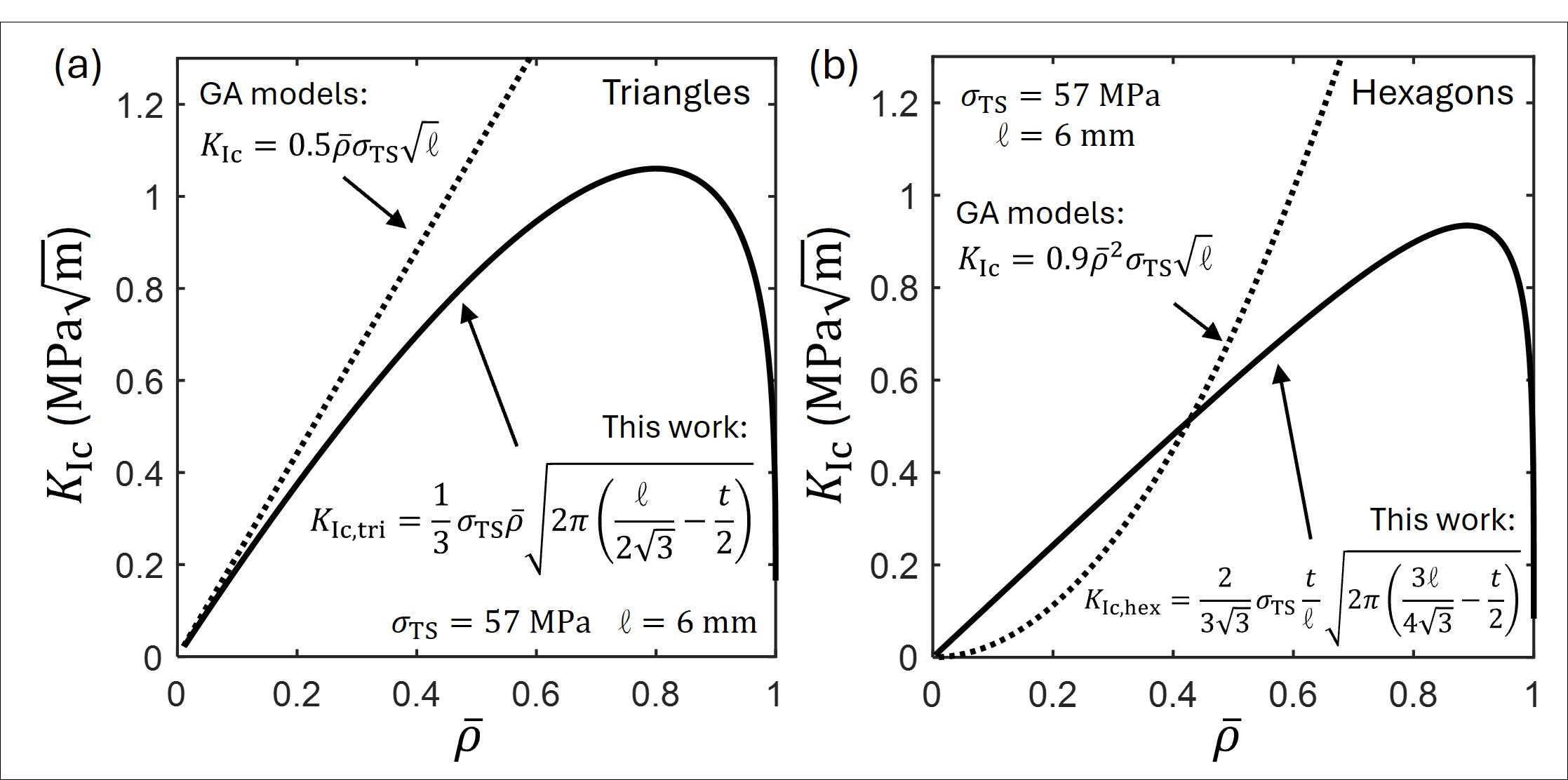}
\caption{Comparison of homogenized, elastic-brittle lattice models to GA models (\autoref{eq:KIclattice}) for (a) triangular and (b) hexagonal lattices.}\label{Fig4}
\end{figure}

The stress at the edge of a circular notch, $\sigma_n$, is given in \autoref{eq:sigTanNotch} and is a function of the notch stress intensity factor which has the general form: $K_{\textnormal{I}}^{\textnormal{n}}=Y\sigma\sqrt{\pi a}$, where $Y$ is some geometry correction factor and $\sigma$ is some applied stress. For the homogenized lattice, $\sigma=\sigma_e$, where $\sigma_e$ is a function of the original applied stress, $\sigma_a$, and the lattice geometry. Grouping $Y\sigma_a\sqrt{\pi a}$ such that the stress intensity factor of the \textit{lattice}, $K_\textnormal{I}$, now depends on the original applied stress, assuming an identical geometry correction factor, $Y$, and taking $\sigma_n=\sigma_\textnormal{TS}$ for a perfectly elastic-brittle lattice material, produces an estimate of fracture toughness for a triangular lattice:
\begin{equation}
    K_{\textnormal{Ic,\textnormal{tri}}}=\frac{1}{3}\sigma_{\textnormal{TS}}\bar{\rho}\sqrt{2\pi \left(\frac{\ell}{2\sqrt{3}}-\frac{t}{2}\right)}
\label{eq:KIcHomogTRI}
\end{equation}
and for a hexagonal lattice:
\begin{equation}
    K_{\textnormal{Ic,\textnormal{hex}}}=\frac{2}{3\sqrt{3}}\sigma_{\textnormal{TS}}\left(\frac{t}{l}\right)\sqrt{2\pi \left(\frac{3\ell}{4\sqrt{3}}-\frac{t}{2}\right)}.
\label{eq:KIcHomogHEX}
\end{equation}
At very low-$\bar{\rho}$ where $t\ll \ell$, \autoref{eq:KIcHomogTRI} converges closely to the GA models as seen in \autoref{Fig4}(a). The stretch-dominated nature of the triangular lattice, in which the struts are dominated by axial stresses, is more conducive to homogenization than the hexagonal lattice. The homogenization approach for hexagons does not account for the large bending stresses present in low-$\bar{\rho}$ hexagonal lattices, whose fracture toughness scales with $\bar{\rho}^2$, but likely better predicts fracture toughness when $\bar{\rho}>0.3$ as bending stresses diminish and the stress distribution is more uniform. When comparing these models to the FE results, they should be truncated at very high-$\bar{\rho}$ when the theoretical notch radius equals the FE fillet radius (which occurs at around $\bar\rho=0.92$ for triangular lattices and $\bar\rho=0.96$ for hexagonal lattices with $\ell=6$ mm and $R=0.5$ mm).

\section{Finite Element Model}\label{femodel}
FE simulations of triangular and hexagonal lattices were performed using Abaqus v. 6.21 (Dassault Systemes, Velizy-Villacoublay, FR), incorporating the methods presented in \autoref{latticeFailureMethods} and \ref{crackGrowthMethod}. The cell size, $\ell$, is held constant at 6 mm while $t$ is varied from 0.18 mm to 2.89 mm to produce relative densities between 0.1 and 0.9 in both triangular and hexagonal lattices. An out-of-plane thickness, $b$, of 2.92 mm is used for all simulations, consistent with the thickness of the experimental specimens. Four relative orientations are investigated in this study ($\theta=0^\circ,10^\circ,20^\circ,30^\circ$) to assess its effect on lattice fracture toughness. The radii of all fillets are held constant at $R=0.5$ mm in the main results to ensure consistent sharpness at each node across all geometries. A separate study on fillet radius (with $R=0.25,0.50,0.75,1.0$ mm) is performed on a subset of lattice geometries to understand its independent effect on the fracture process. Each lattice is comprised of 25 $\times$ 25 unit cells to produce a proper K-field at the crack tip. The size of the specimen was verified through a convergence study by increasing the number of cells until the measured toughness converged within 3$\%$. A linear elastic material model is used in all FE modeling, with PMMA elastic properties. PMMA tensile strength, $\sigma_{\textnormal{TS}}$, and Young's Modulus, $E_\textnormal{b}$, were determined experimentally from five dogbone specimens (ASTM D638 Type IV) and five rectangular strips (with nominal dimensions: 50 mm $\times$ 10 mm $\times$ 3 mm), respectively. When loaded at a strain rate consistent with experimental fracture tests (fracture tests are loaded at a displacement rate of 1 mm/min), $E_\textnormal{b}=2300\pm200$ MPa, and $\sigma_{\textnormal{TS}}=57\pm3$ MPa. The Poisson's ratio of PMMA, $\nu$, was taken as $0.35$ \parencite{maccagno1989fracture}. Finally, PMMA plane-strain fracture toughness was determined to be $K_\textnormal{Ic,b}=1.30\pm 0.16$ MPa$\sqrt{\textnormal{m}}$ from five compact tension (CT) specimens using the method presented by Bower \parencite{bower2009applied}.

\begin{figure}[h]
\centering
\includegraphics[width=1\textwidth,trim={0.2cm 0.2cm 0.2cm 0.2cm},clip]{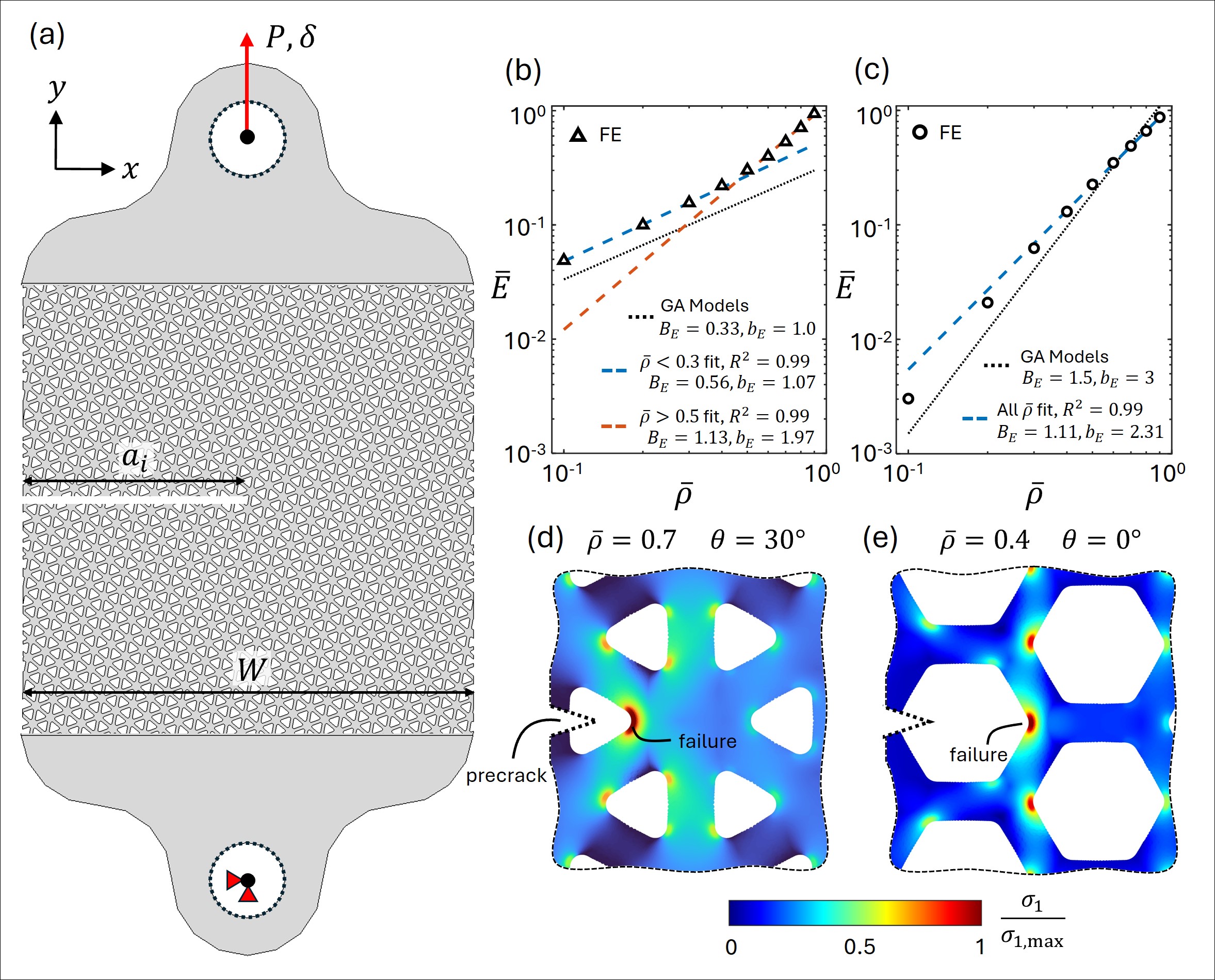}
\caption{Finite element lattice model. (a) Overall specimen geometry of the modified SENT fracture specimen to enable pin-loading. Effective lattice Young's Moduli for (b) triangular and (c) hexagonal lattices across a wide range of relative densities. Trends are fit using a power law to compare directly to GA models for lattices at low relative density. Two regimes in relative density are used to fit the triangular lattice results, and only one regime is necessary for hexagons. The maximum principal stress distribution ahead of the precrack in a (d) triangular and (e) hexagonal lattice.}\label{Fig5}
\end{figure}

The overall specimen geometry used in the FE analysis is designed to facilitate simple experimental testing. The geometry includes holes for pin mounting on the top and bottom of the lattice, as shown in \autoref{Fig5}(a), which also helps improve alignment during testing. In the FE model, the perimeter of both holes are constrained to their center points using a rigid-body constraint to mimic a pin load. The center of the bottom hole is constrained in the $x$- and $y$-directions, while the top hole is displaced in +$y$ and constrained in $x$. Additionally, both holes are free to rotate during loading.

All models were meshed with 6-node quadratic plane-strain triangular elements (CPE6 elements), and the mesh was significantly refined around each fillet to capture the large stress gradients there. An element size of 0.09 mm at each fillet for each lattice geometry was sufficient to produce less than a 1$\%$ deviation in peak stress with increasing refinement. Less refinement was necessary in the strut regions away from the fillets, so an element size of $t/5$ was used for all geometries. Using this seeding, each FE model consisted of a mesh with approximately $10^6$ elements. 

Effective moduli for each lattice geometry are calculated through separate FE simulations on an uncracked specimen. The measured lattice effective moduli, $E_f$, are normalized by the base material modulus, $E_{\textnormal{b}}=2.3$ GPa, to yield a non-dimensional $\bar{E}$ (and the results are plotted in (\autoref{Fig5}(b-c)). A power law with the form: $\bar{E}=E_f/E_{\textnormal{b}}=B_E\bar{\rho}^{b_E}$ is fit to the results using nonlinear least squares for different ranges of relative density and compared to the GA models, where $B_E$ and $b_E$ are fitting coefficients. Slight differences between the GA models at low relative densities are attributed to the fillet radii in the FE model, which increase stiffness. Two regimes appear for triangular lattices with a transition point around $\bar{\rho}=0.4$. At high-$\bar{\rho}$, the effective moduli of both triangular and hexagonal lattices are expected to scale with relative density in the same way. However, at low densities, triangular lattices carry mostly axial stresses, resulting in increased stiffness compared to the hexagonal lattice. The hexagonal lattice's moduli, on the other hand, follows a single trend with relative density. The values for $B_E$ and $b_E$ can be found in \autoref{Fig5}(b-c) along with the R$^2$ of each fit.

Lattice toughness is extracted for each geometry using the methods described in \autoref{latticeFailureMethods} and \autoref{crackGrowthMethod}. A custom Python script extracts the $\sigma_1$ stress-field, load, displacement, and total strain energy to implement the failure criteria. Failure initiation in each lattice occurs according to either the elastic-brittle or quasi-brittle failure criteria, and a crack grows through the unit cell according to the MERR criterion. In the elastic-brittle criterion, failure occurs whenever $\sigma_{1,\textnormal{max}}=\sigma_\textnormal{TS}$, while the quasi-brittle criterion looks at a region ahead of the largest stress concentration to determine failure. The quasi-brittle failure criterion is able to account for localized plasticity using just the linear-elastic stress distribution. Following \autoref{eq:rc} and \autoref{eq:dc} and using the average fracture toughness and $\sigma_\textnormal{TS}$ values found for PMMA, the critical distances are calculated as: $r_c=$0.082 mm and $d_c=$ 0.331 mm. Representative stress distributions ahead of the largest stress concentration in high-$\bar{\rho}$ triangular and hexagonal lattices are shown in \autoref{Fig5}(d-e) and may be used to inform either the PS or MS criterion as described in \autoref{latticeFailureMethods}. Because the stress concentration is not always symmetric about the midplane, the PS or MS criteria are evaluated along the line of the crack path direction found by the MERR criterion. The FE elastic stress distribution ahead of the point of maximum stress is compared to the theoretical stress distribution defined by notch fracture theory in \ref{app1} to confirm its validity.

The MERR criterion is simultaneously employed to determine how a crack grows through each unit cell. Starting at the point of maximum stress, lines are drawn along the shortest path to each adjacent cell as shown in \autoref{Fig2}(b). Cuts along each path are added to the FE model individually, and a simulation is conducted for each. The total elastic strain energy is measured before and after a cut is made, and the difference is normalized by the crack area made from this cut, $A_i=bc_i$, where $c_i$ is the length of the cut path as seen in \autoref{Fig2}(b). The path that maximizes $G=-\Delta U/ \Delta A$ is identified as the crack. The FE model is finally updated with the correct crack path, and the process is repeated at least six times to obtain multiple stable ruptures. This corresponds to six stable points along the load-displacement curve that are used to calculate an average $G_c$. The MERR criterion is validated in \ref{app2} where separate extended finite element method (XFEM) simulations are compared to the crack paths predicted by MERR, showing great agreement especially for $\bar{\rho}<0.9$.

The compliance method is used to calculate lattice toughness from the FE results as described in \autoref{angComMethod}. First, the fitting coefficients, $D_1$ through $D_6$, in \autoref{eq:CvsAcurve} are found by constructing a FE model with the identical overall specimen geometry as in \autoref{Fig5}(a), except the lattice geometry is replaced by a homogeneous material with an effective modulus $E_f$ and precrack with length $a_i$. Specimen width, $W$, is 147 mm and 135 mm for the triangular and hexagonal lattice specimens, respectively. The compliance of the added material above and below the lattice (grip compliance) is calculated by removing the center test material from the model and simulating the response of just the top and bottom pin grips. Grip compliance is found to be $C_g=0.236$ ($\mu$m/N), assuming a modulus, $E_g=2300$ MPa, and thickness, $b=2.92$ mm. The compliance of the center specimen, $C_f$, is found by setting $E_f\ll E_g$ to minimize the compliance contribution of the grips, then simulating the system response as a function of total projected crack length, $a_{p,t}$. The assumption that the deflected crack specimen compliance is approximated by the projected crack specimen compliance is validated from FE studies as described in \ref{app3}. $D_1$ through $D_6$ are found by fitting \autoref{eq:CvsAcurve} to the FE $C_f$ vs. $a_{p,t}$ curve via nonlinear least squares ($D_1,...,D_6=$ 2789, -6500, 6039, -2765, 622.5, and -54.64 $\textnormal{mm}^{-1}$ for $b=2.92$ mm) as described in detail in \ref{app3}. Then, the total specimen compliance, $C$, is calculated, where: $C=C_f+C_g$.

For lattice FE specimens, the specimen compliance is used to find $a_{p,t}$ and $a_{\theta,t}$, using \autoref{eq:CvsAcurve} and calculating $a_i$ and $\theta_d$ from the full FE crack path. Finally, $G_c$ can be calculated from FE measurements of peak loads, $P_c$, and using the equation described in \autoref{angComMethod}, $G=f(P,b,dC/da_{\theta,t})$. The results are converted to a fracture toughness, $K_\textnormal{Ic}$, using the Irwin relation, and a mean and standard deviation are calculated across all rupture events.

\section{Experimental Methods}\label{expMethods}
\begin{figure}[!htb]
\centering
\includegraphics[width=1\textwidth,trim={0.2cm 0.1cm 0.2cm 0.2cm},clip]{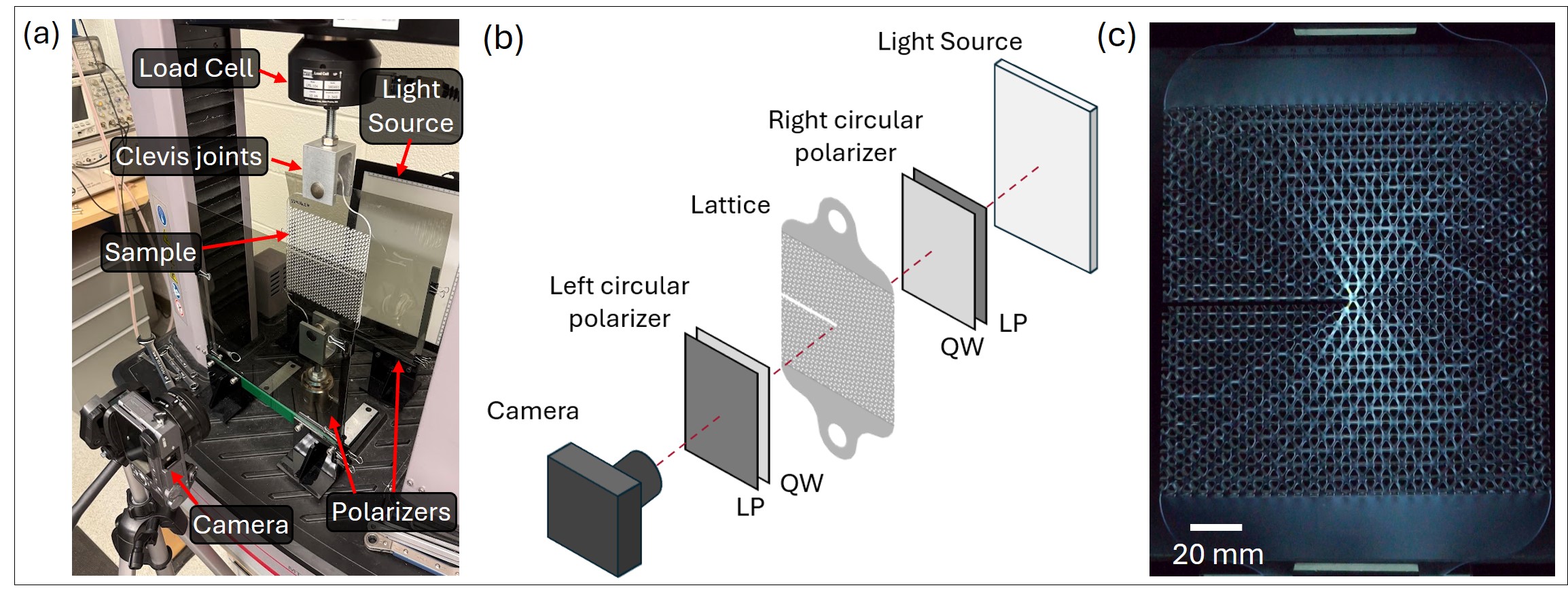}
\caption{Experimental design. (a) Experimental set-up in a uniaxial tensile testing machine, including a photoelasticity set-up. (b) Photoelasticity set-up schematic showing a lattice sample between left and right circular polarizers. (c) Example photoelastic image from a triangular lattice fracture test where regions of high light intensity correspond to regions of high internal stress.}\label{Fig6}
\end{figure}

Lattice specimens were laser-cut from homogeneous PMMA sheets (optically transparent, and with sheet thickness, $b=2.92$ mm) using the overall specimen geometry shown in \autoref{Fig5}(a). Samples were then annealed at 90 $^\circ$C for 4 hours to reduce internal stresses that form during laser-cutting \parencite{fulco2024enhancing}. Triangular and hexagonal lattices with three relative densities (spanning 0.3 to 0.8) and two relative orientations ($0^\circ$ and $30^\circ$) were fabricated to capture the range of fracture behavior. Three samples for each geometry were fabricated to measure the average response of each lattice geometry, as any geometric or material imperfection may affect the fracture process. Each sample was mounted to a uniaxial testing machine (Criterion Model 43, MTS) using pins and clevis joints as seen in \autoref{Fig6}(a). Each test was run continuously at a displacement rate of 1 mm/min until the crack grew through the entire sample. Load and displacement were measured throughout the test to compare with the FE results. The displacement data was calibrated for machine compliance and was offset to account for initial slack in the system.

As PMMA is birefringent, the internal stress field could be visualized during testing using polarized light, an effect known as photoelasticity. The photoelasticity set-up can be seen in \autoref{Fig6}(b), where the lattice is positioned between two sets of polarizers, each having a linear (LP) and quarter-wave (QW) polarizer \parencite{daniels2017photoelastic}. When light travels from the back of the setup, passes through the polarizer-sample system, and reaches the viewer, the intensity of light seen in the sample is proportional to the difference in the maximum in-plane principal stresses. An example image from one PMMA lattice test is shown in \autoref{Fig6}(c) at the point right before failure. We use photoelasticity to record the stress field throughout each test and to track the position and growth of the macroscopic crack through each sample.

\section{Results and Discussion}
\subsection{Elastic-Brittle Lattices}

\begin{figure}[!htb]
\centering
\includegraphics[width=1\textwidth,trim={0.2cm 0.4cm 0.2cm 0.2cm},clip]{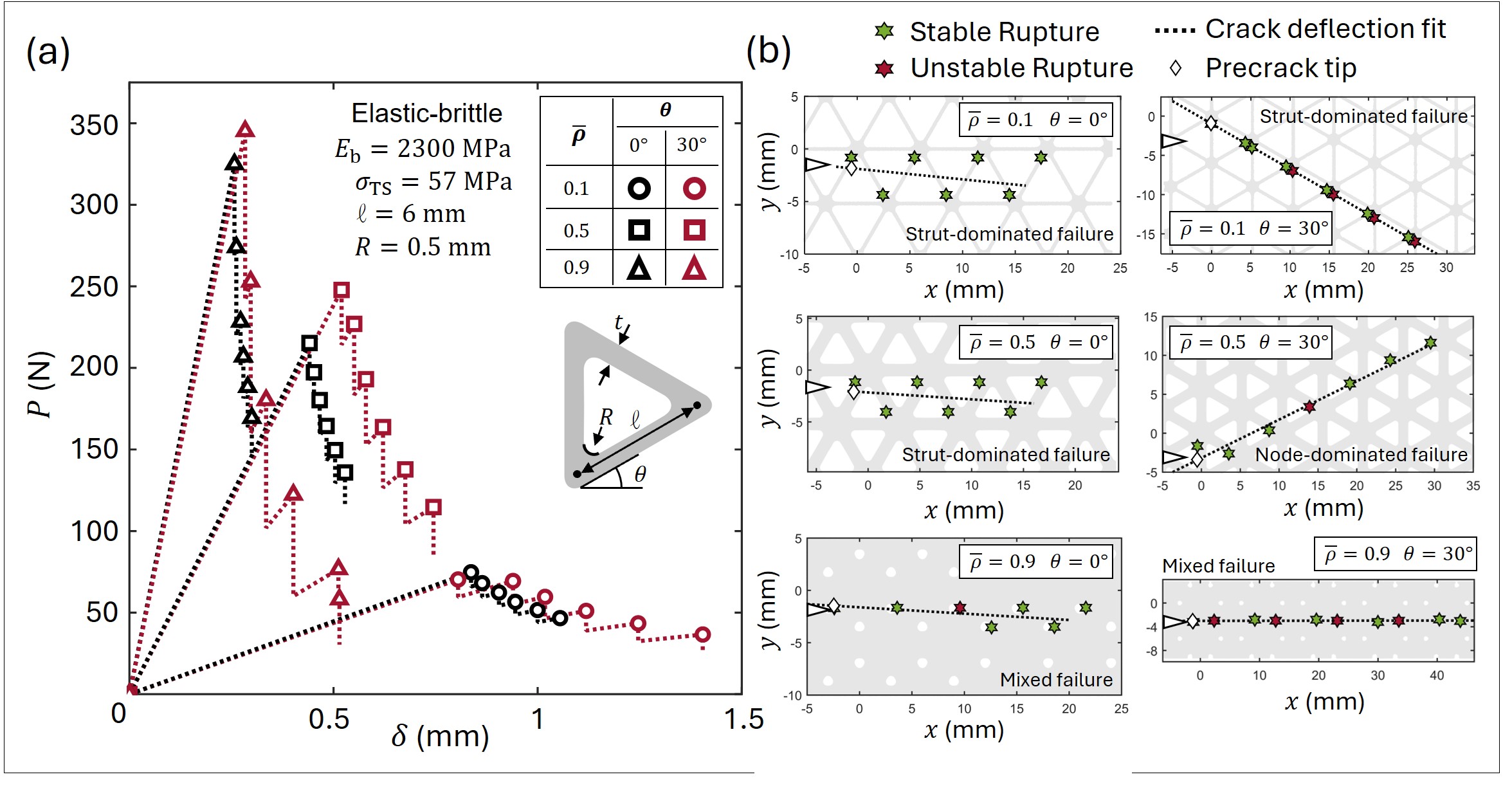}
\caption{Triangular lattice FE results. (a) The load-displacement curves for a subset of triangular lattice geometries varying both relative density and relative orientation. Markers indicate the peak load before a rupture event. (b) Crack paths and rupture locations in the corresponding subset of triangular lattices.}\label{Fig7}
\end{figure}

\begin{figure}[!htb]
\centering
\includegraphics[width=1\textwidth,trim={0.2cm 0.3cm 0.2cm 0.2cm},clip]{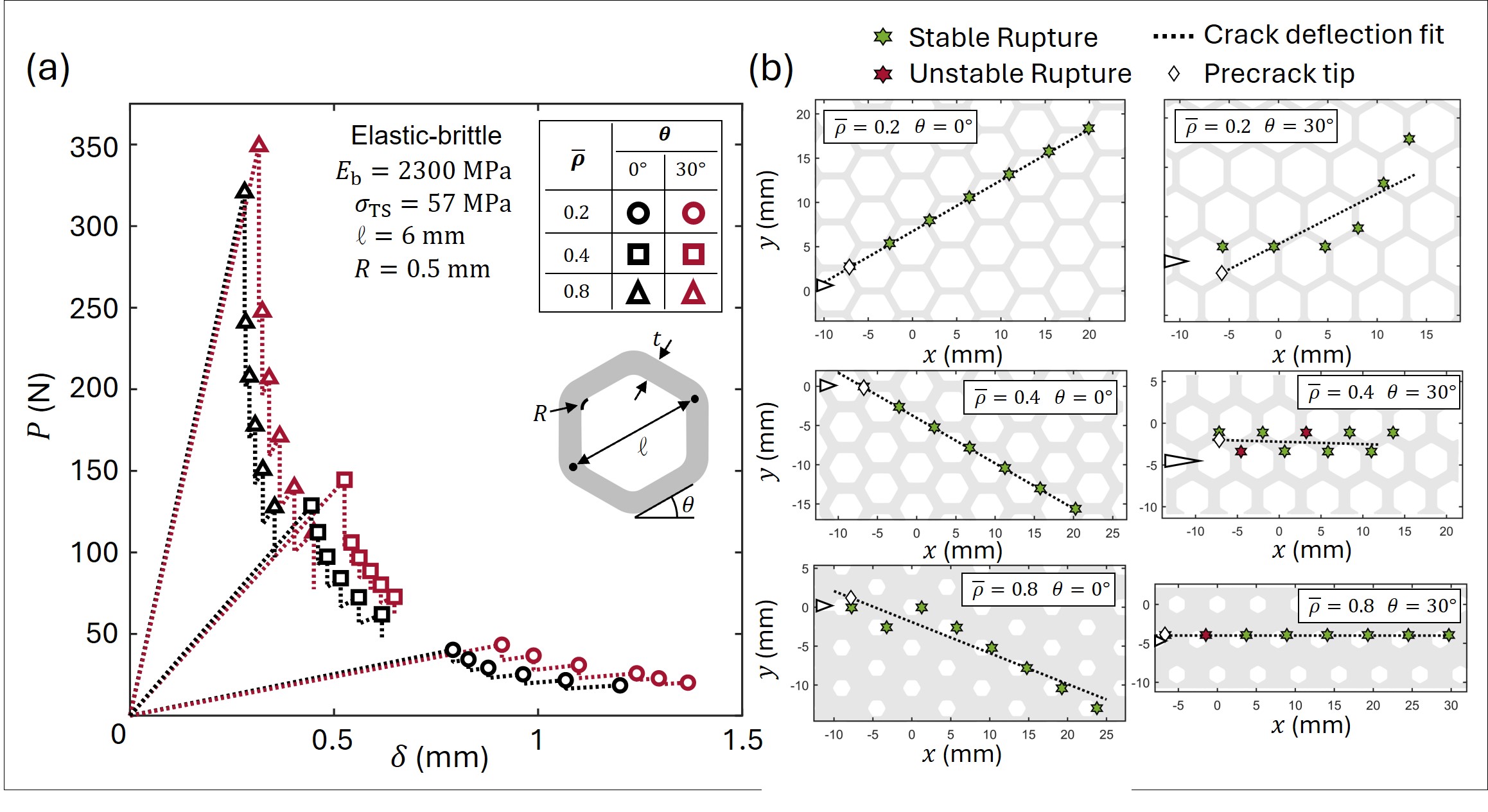}
\caption{Hexagonal lattice FE results. (a) The load-displacement curves for a subset of hexagonal lattice geometries varying both relative density and relative orientation. Markers indicate the peak load before a rupture event. (b) Crack paths and rupture locations in the corresponding subset of hexagonal lattices.}\label{Fig8}
\end{figure}

Representative FE results for elastic-brittle triangular and hexagonal lattices are shown in \autoref{Fig7} and \autoref{Fig8}, respectively. The force-displacement curves and the associated crack paths for three relative densities and two relative orientations are shown. Each lattice geometry exhibits an initially linear response before failure, followed by progressive fracture of cells within the lattice that correspond to sudden drops in load (forming the damage curve). Stable fracture is observed across all lattice geometries, enabling calculation of a steady-state fracture toughness. Lattices with the same relative density have nearly identical initial stiffness regardless of changes to relative orientation, suggesting an isotropic elastic response. However, the damage curve is elevated in the $\theta=30^\circ$ case for both triangular and hexagonal lattices, indicating increased toughness.

The crack paths through these lattices provide insight into their fracture behavior. The direction a crack will travel depends on the local geometry of the crack tip, the loading configuration, and the history of damage. Triangular lattices at $\bar{\rho}=0.1,0.5$ have cracks that run along the row of cells that are parallel to the lattice orientation, while at higher relative densities, this effect diminishes as the crack is not as easily pulled along this weakened path of angled struts. Instead, the crack simply grows straight at high-$\bar{\rho}$. Crack paths in hexagonal lattices are less affected by relative density than in triangular lattices, but remain dependent on relative orientation where straight cracks are expected in the $\theta=30^\circ$ case and deflected cracks are present in all other orientations. The angle of the crack will have an impact on the local stress distribution -- more deflected cracks will introduce greater mixed-mode effects. In addition to the crack angle, the local geometry at the crack tip will dictate how stress is distributed there and influence when failure occurs.

\begin{figure}[!htb]
\centering
\includegraphics[width=1\textwidth,trim={0.15cm 0.1cm 0.3cm 0.2cm},clip]{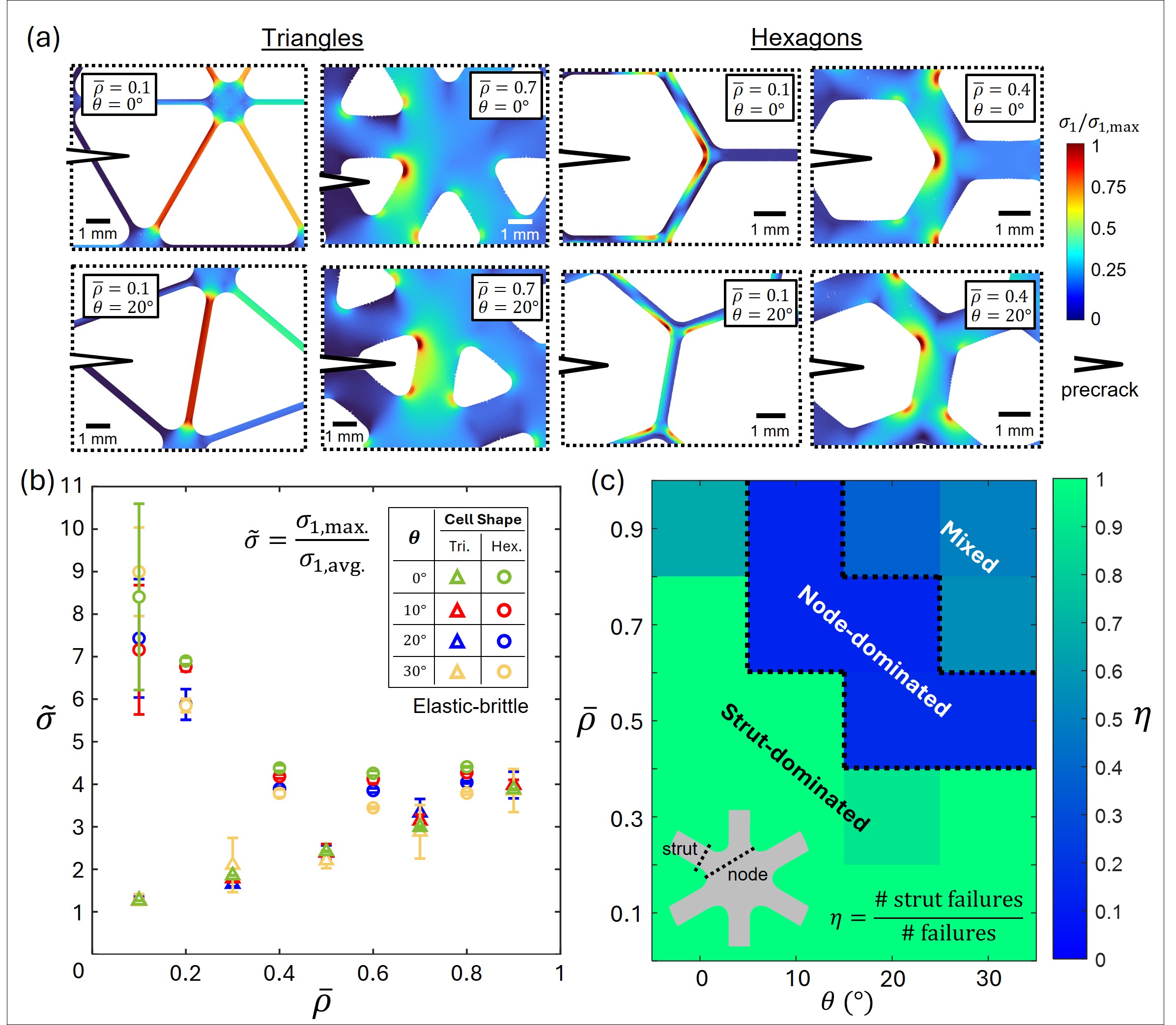}
\caption{Results from the FE model. (a) The distribution of stress in triangular and hexagonal lattices for different relative densities and orientations. (b) The magnitude of the stress concentration present in the leading strut of the lattice, captured by normalizing the peak strut stress with the average strut stress. Error bars capture the standard deviation across multiple measurements of normalized stress during crack growth. (c) Rupture paths in the triangular lattice cell as a function of relative orientation and density.}\label{Fig9}
\end{figure}

Stress distributions at the crack tip are shown in \autoref{Fig9}(a) for triangular and hexagonal lattices of different $\bar{\rho}$ and $\theta$. At $\bar{\rho}=0.1$, a nearly uniform uniaxial stress develops in triangular lattices. Failure can be approximated knowing this uniaxial stress, as is assumed in the GA models. Similarly, a combination of axial and bending stresses develops in hexagonal lattices, depending on the relative orientation $\theta$, and these stresses can be approximated using beam theory. At higher $\bar{\rho}$, however, stress concentrations develop near the nodes of both triangular and hexagonal lattices and influence the fracture process. While the stress distribution is primarily controlled by relative density (a function of the aspect ratio of the strut, where $t/\ell \propto \bar{\rho}$), it is also affected by the local geometry at the crack tip, as seen in \autoref{Fig9}(a). Crack tip geometry depends on the relative orientation of the lattice, $\theta$, the cell shape (triangular or hexagonal), and the fillet radius, $R$. 

To observe the influence of stress concentrations on failure, we approximate the magnitude of the largest stress concentration by looking at the ratio of maximum to average $\sigma_1$ in the strut carrying the highest stress, $\tilde\sigma=\sigma_{1,\textnormal{max}}/\sigma_{1,\textnormal{avg}}$, as shown in \autoref{Fig9}(b). Error bars capture the standard deviation across multiple rupture events for each lattice geometry. The average stress, $\sigma_{1,\textnormal{avg}}$, is found by taking an average of the stresses distributed in a circular region located at the center of the strut with a diameter equal to the strut thickness. In low-$\bar{\rho}$ triangular lattices, $\tilde\sigma$ is very close to 1 as expected, confirming a nearly uniform stress distribution. As relative density increases, $\tilde\sigma$ increases to about 4, where large stress concentrations become more prevalent. While the magnitude of the stress concentration increases with relative density, so does the volume of material that carries the far-field load, thereby increasing fracture toughness. Thus, increasing relative density while minimizing $\tilde\sigma$ will produce the toughest lattice. Hexagonal lattices have the opposite behavior -- at low $\bar{\rho}$, large bending stresses dominate the stress distribution and $\tilde{\sigma}\approx8$. At higher $\bar{\rho}$, struts begin to shear instead of bend, leading to a more uniform distribution across the lattice, and $\tilde\sigma$ decreases to about 4, a similar value as seen in triangles. Additionally, changing the relative orientation yields slightly different values of $\tilde\sigma$ as the size and shape of the stress concentration changes, as shown in \autoref{Fig9}(a). Lattices with $\theta=30^\circ$ tend to have the smallest stress concentrations for a given $\bar{\rho}$ due to the presence of vertical struts at the crack tip. In this case, stress is distributed more equally between the adjacent nodes, reducing the magnitude of both. This is true for hexagonal lattices regardless of the crack tip location. However, the macroscopic crack may also travel into very sharp features in triangular lattices oriented at $\theta=30^\circ$, leading to a larger variance in $\tilde\sigma$ and toughness. 

\begin{figure}[!tb]
\centering
\includegraphics[width=1\textwidth,trim={0.2cm 0.2cm 0.6cm 0.2cm},clip]{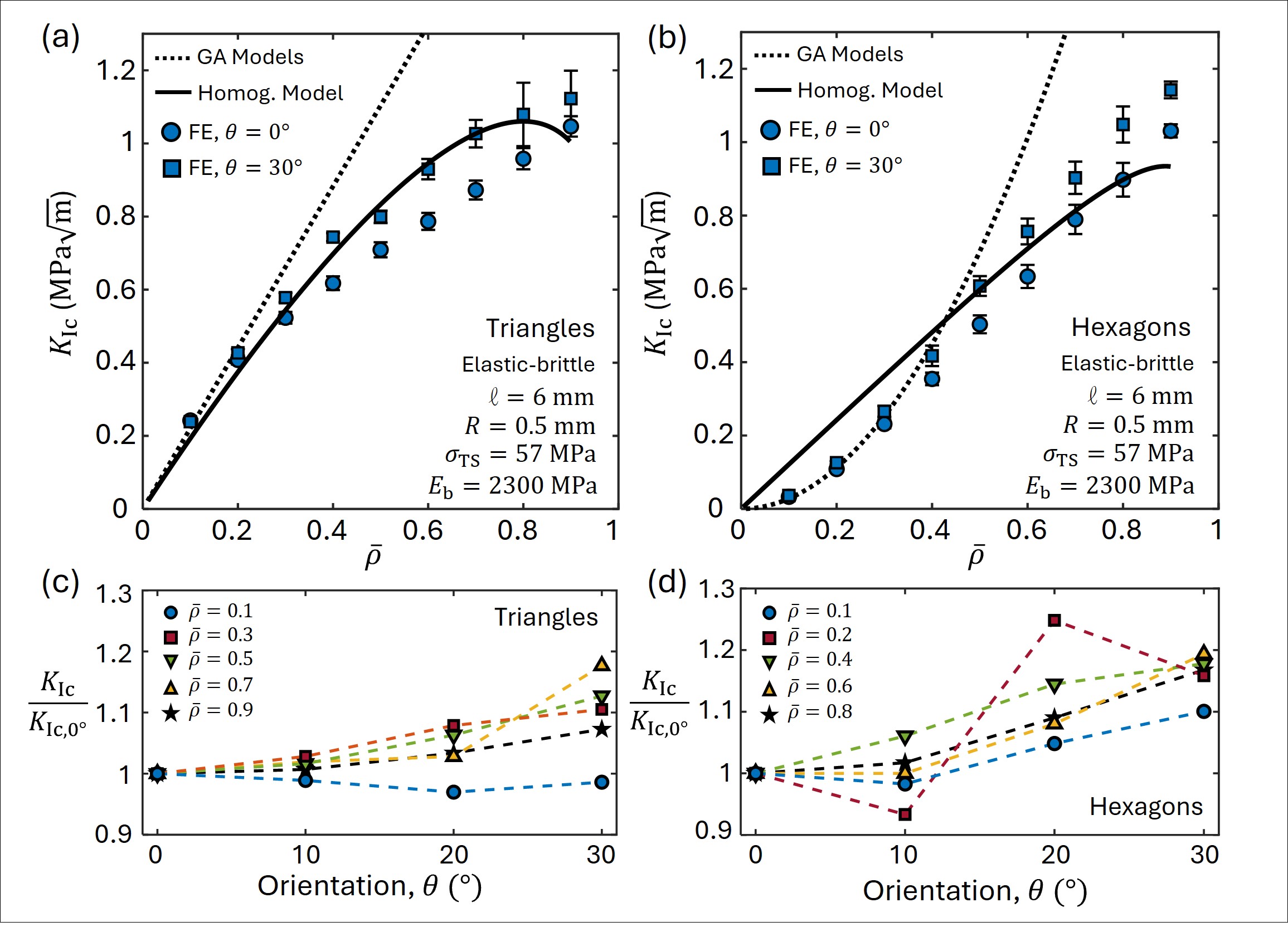}
\caption{Fracture toughness results for elastic-brittle lattices. Fracture toughness as a function of relative density for (a) triangular and (b) hexagonal lattices, comparing the FE results to GA models and the developed homogenized lattice models. Error bars show the standard deviation across multiple measurements of fracture toughness during crack growth (i.e., toughness is calculated at each rupture event). (c) Triangular and (d) hexagonal lattice FE fracture toughness results are then normalized by the 0$^\circ$ case and plotted against orientation to observe fracture anisotropy.}\label{Fig10}
\end{figure}

The direction a crack will grow through the cell of a triangular lattice becomes unclear at high-$\bar{\rho}$ (when $\bar{\rho}>0.4$, generally) as a crack originating at the maximum stress concentration near a node is no longer limited to strut-based failure. As $\bar{\rho}$ increases, the distance the crack must travel for strut versus nodal failure becomes comparable, rendering nodal failure more energetically favorable in certain configurations. The crack will always follow the most energetically favorable path -- in some cases, it will grow along the lattice orientation as the struts along this path typically carry the largest stresses, while in others, the crack simply grows straight (parallel to the $x$-axis), consistent with mode-I fracture. The transition in behavior from strut-dominated to node-dominated to mixed is summarized in \autoref{Fig9}(c), where strut failure is represented by: $\eta=\textnormal{\# strut failures / \# failures}=1$. At $\theta=0^\circ$ and for all $\bar{\rho}$, the weakest path in the lattice is parallel to the $x$-axis, so the crack consistently grows straight and only ruptures through struts. At the right combination of $\bar{\rho}$ and $\theta$, there is no clear preference for the crack to grow along the lattice orientation or grow straight. As a result, the macroscopic crack deflects along the lattice orientation, but each failure event occurs through the node rather than the strut as it is along a more straight path (this behavior can be observed in \autoref{Fig7}(b) for $\bar{\rho}=0.5$ and $\theta=30^\circ$). Finally, at high $\bar{\rho}$ and $\theta$, the most energetically favorable path is straight, causing a mix of both strut and nodal failure. Within the mixed regime, a greater variance in fracture toughness is expected.

\begin{figure}[!htb]
\centering
\includegraphics[width=1\textwidth,trim={0.3cm 0.2cm 1.4cm 0.2cm},clip]{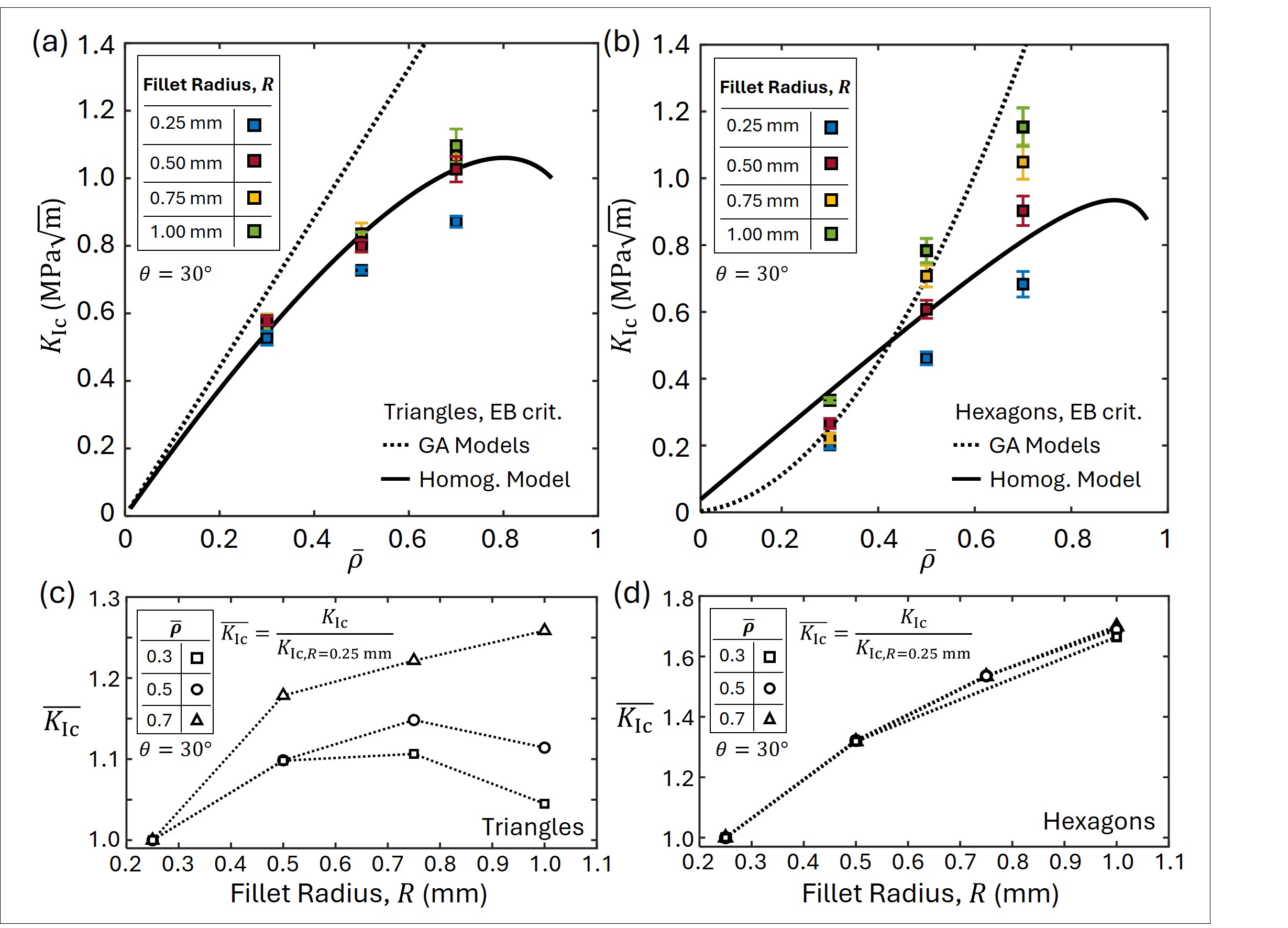}
\caption{Effect of fillet radius on elastic-brittle lattice fracture toughness. FE fracture toughness results versus relative density for (a) triangular and (b) hexagonal lattices and varying fillet radius. FE results are compared to GA models and the developed homogenized lattice models. Error bars capture multiple measurements of fracture toughness during crack growth. (c) Triangular and (d) hexagonal lattice FE fracture toughness results are then normalized by the smallest fillet radius case and plotted against fillet radius.}\label{Fig11}
\end{figure}

Fracture toughness, $K_{\textnormal{Ic}}$, is extracted for all lattice geometries, assuming elastic-brittle failure and plotted as a function of relative density in \autoref{Fig10}(a-b) to compare to GA models and our homogenized lattice model. Fracture toughness increases with relative density, but diverges from the GA models for $\bar\rho>0.4$. At low-$\bar{\rho}$, the FE results converge to the GA models as is anticipated from the $\tilde\sigma$ results. For $\bar{\rho}>0.4$, the homogenized lattice model better predicts failure in both triangular and hexagonal lattices, capturing the magnitude of $\tilde\sigma$ at the crack tip. The triangular lattice model appears to better match the $\theta=30^\circ$ case across all densities, likely due to the symmetry present in the $30^\circ$ lattice. Meanwhile, the hexagonal lattice model captures the scaling between $K_{\textnormal{Ic}}$ and $\bar{\rho}$ for $\bar{\rho}>0.4$ while GA models better predict failure at low-$\bar{\rho}$. The FE fracture toughness results appear to converge at very high densities to the fracture toughness of the solid elastic-brittle material ($\bar\rho=1$) with a notch radius equal to the fillet radius, $R_n=R=0.5$ mm, of about 1.06 MPa$\sqrt{\textnormal{m}}$. The FE study is restricted by the size of the fillet radius -- fracture toughness values would likely significantly decrease, similar to the homogenized lattice model, when $\bar\rho>0.9$ if an even sharper crack ($R_n<0.5$ mm) were introduced. Regardless, the homogenized lattice model performs well across a wide range of relative densities despite zero fitting to the FE results and using only the independently measured material properties of the base material and the original lattice geometry. The model would be improved by introducing fitting parameters, as done in GA models, but this is beyond the scope of this work.

The separation in fracture toughness results between $\theta=0^\circ$ and $\theta=30^\circ$ geometries is small at low-$\bar{\rho}$ as expected. These results are corroborated by similar findings in the literature for low-$\bar{\rho}$ triangular and hexagonal lattices \parencite{chen1998fracture,lipperman2007fracture,gu2018experimental,bijaya2023multiscale}. However, as relative density increases, the difference in fracture toughness with orientation grows. To better observe this increase in fracture anisotropy, we plot fracture toughness normalized by the $\theta=0^\circ$ case in \autoref{Fig10}(c-d). At $\bar{\rho}=0.7$, fracture anisotropy is maximum and normalized fracture toughness, $K_\textnormal{Ic}/K_{\textnormal{Ic,}0^\circ}$, exceeds 1.2. This increase in anisotropy is attributed to the stress field around the crack tip being more sensitive to changes in lattice orientation at higher relative densities, as well as differences in the macroscopic crack paths. Finally, when $\bar{\rho}>0.8$, generally, the crack does not deflect as significantly, and the crack tip feature becomes more symmetric, leading to lower fracture anisotropy. Overall, the magnitude of fracture anisotropy is relatively small (below $25\%$), but the $\theta=30^\circ$ case consistently has the highest fracture toughness of all tested orientations regardless of the relative density.

As noted previously, the local crack tip geometry will have a large effect on the magnitude and shape of the dominant stress concentration, especially at high-$\bar\rho$. The fillet radius at each corner in the lattice will contribute to crack sharpness and affect failure. We extract the fracture toughness of lattices with different fillet radii ($R=0.25,0.5,0.75,1$ mm) for a subset of relative densities and orientations as shown in \autoref{Fig11}. Adding fillets to the lattice geometry will add material to the system and increase relative density; however, nominal relative densities (assuming no fillets) are used here for comparison between lattice geometries. Triangular lattice fracture toughness appears to plateau as $R$ increases and converges close to our model across all densities. However, in the $\bar{\rho}=0.3,0.5$ cases, continuing to increase fillet radius actually leads to a slight \textit{decrease} in fracture toughness. For the triangular lattice, increasing $R$ leads to two competing effects: 1) the crack tip geometry at the dominant stress concentration becomes more blunt, so stress is less localized, and 2) the length of the strut is effectively shortened as $R$ increases, leading to more localized stress. This competition can explain the discrepancy in trends at different relative densities, as the significance of either mechanism changes with relative density. Because the stress distribution in low-$\bar{\rho}$ hexagonal lattices is already very concentrated near the node, increasing $R$ leads to a monotonic increase in fracture toughness. At high-$\bar{\rho}$, hexagonal lattices continue to benefit from the increase in $R$, similar to triangles. While the homogenized lattice model captures the general scaling behavior of hexagonal lattices for $\bar{\rho}>0.4$, it cannot account for the complicated stress distribution within struts at the crack tip.



\subsection{Quasi-brittle Lattices}

To determine failure in a quasi-brittle lattice, the FE elastic stress distribution ahead of the point of maximum stress is evaluated using quasi-brittle notch fracture theory. The mean stress (MS) criterion is used in all cases except when the lattice geometry confines stress, and the elastic stress distribution is unable to decay. The peak stress (PS) criterion is used when the critical distance defined in the MS criterion is comparable to the strut thickness (occurring around $\bar\rho=0.25$).

\begin{figure}[!htb]
\centering
\includegraphics[width=1\textwidth,trim={0.2cm 0.2cm 0.2cm 0.2cm},clip]{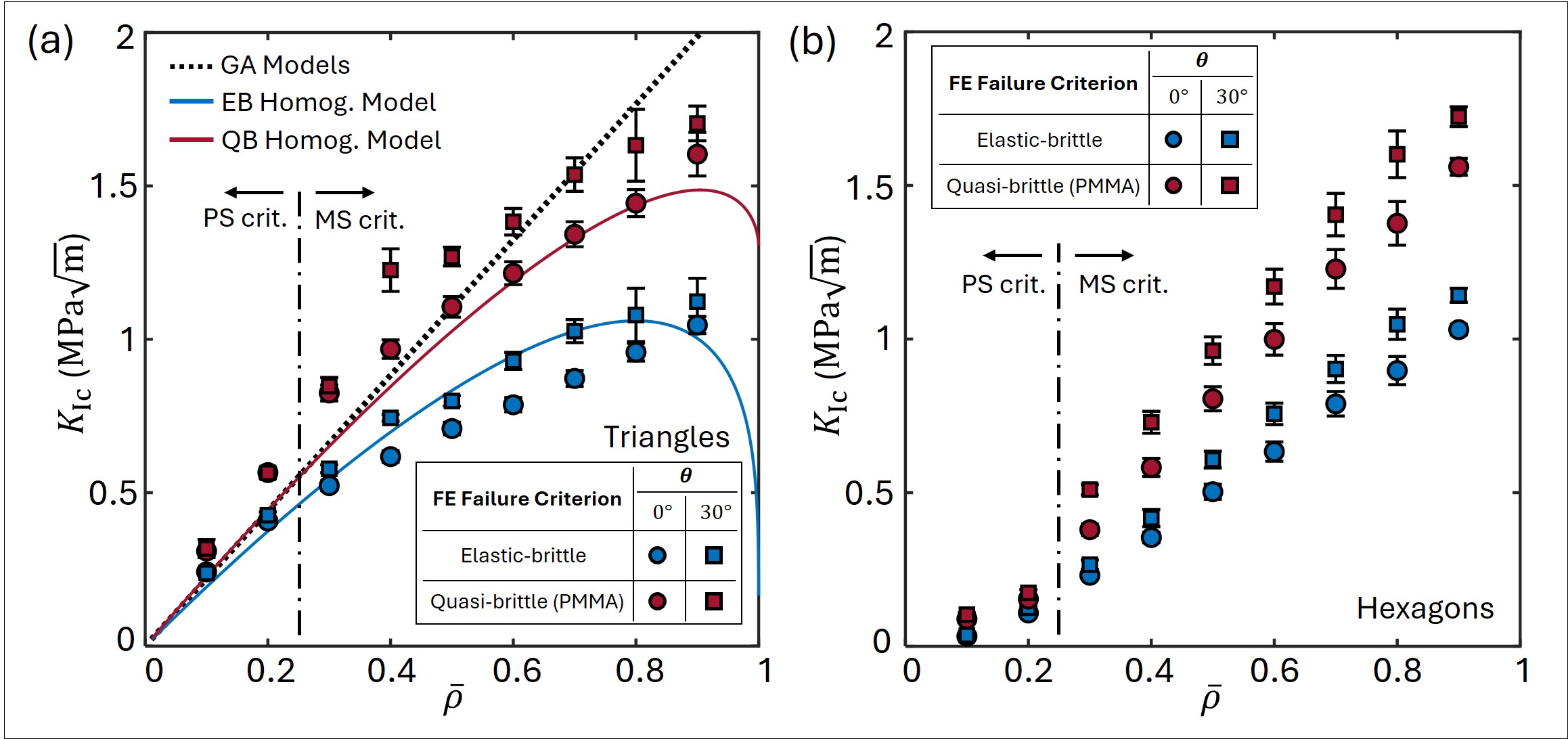}
\caption{Fracture toughness results for quasi-brittle lattices.  The quasi-brittle results (red) are compared to the elastic-brittle results (blue) for (a) triangular and (b) hexagonal lattices. EB and QB homogenized lattice models and GA model curves are added to the triangular results. The mean stress (MS) criterion is used for the FE results in cases where the elastic stress distribution is not affected by the lattice geometry. The peak stress (PS) criterion is used at low relative densities. Error bars capture multiple measurements of fracture toughness during crack growth.}\label{Fig12}
\end{figure}

Simulations of lattices with the quasi-brittle failure criterion showed an increase in fracture toughness of about 1.5-1.6$\times$ compared to the elastic-brittle lattices across almost all geometries, as shown in \autoref{Fig12}. This increase in toughness trends with the magnitude of the dominant stress concentration shown in \autoref{Fig9}(b) -- low-$\bar{\rho}$ hexagonal lattices and high-$\bar{\rho}$ triangular lattice receive the greatest benefit from the quasi-brittle criterion. Similar trends with relative density and relative orientation are observed in quasi-brittle triangular and hexagonal lattices, as in elastic-brittle lattices. 

An alternative derivation of the homogenized lattice model to account for quasi-brittle material behavior is presented in \ref{app5} and plotted in \autoref{Fig12}(a) for the triangular lattice. The model predicts a peak fracture toughness at around $\bar{\rho}=0.9$ and converges to the fracture toughness of PMMA ($K_\textnormal{Ic,b}=1.30$ MPa$\sqrt{\textnormal{m}}$) at $\bar{\rho}=1$ as expected. The model slightly under-predicts the QB FE results, but does a better job of capturing the nonlinear scaling with $\bar{\rho}$ compared to GA models. A comparison to the homogenized lattice models is only performed for the stretch-dominated triangular lattices, as their internal stress state is more appropriate for homogenization, and failure is well-approximated by such models. We expect this model to be valid for other stretch-dominated structures, provided that the appropriate modifications are made to $\sigma_e$ and $R_n$. Further analysis is required to model the fracture process in hexagonal and other bending-dominated lattices.

Interestingly, the quasi-brittle results show that for $\bar{\rho}>0.6$, the lattice fracture toughness exceeds that of homogeneous PMMA. This toughening has been observed in studies of porous media and suggests that materials can be made tougher and lighter by incorporating internal geometry, but at the cost of strength \parencite{yang2002microstructure,bijaya2023multiscale}. However, this will not hold true for all lattices -- the limits of toughness enhancement and designing more efficient lattice structures are explored in \autoref{QBtheoryResults} using the QB homogenized lattice model. 

\subsection{Designing Mass-Efficient Materials}\label{QBtheoryResults}

The presence of a peak in the QB homogenized lattice model as seen in \autoref{Fig12}(a) suggests that some lattice materials can have a higher fracture toughness than their base material, enabling tougher materials at a reduced mass. However, this toughness peak will shift or diminish as the lattice geometry and the base material properties change. The behavior of the $K_\textnormal{Ic}$ vs. $\bar\rho$ curve is studied using the QB homogenized lattice model where two critical relative densities are explored as a function of $\ell/d_c$ (the relative lengthscale between the lattice cell size and the QB fracture lengthscale). 

\begin{figure}[!htb]
\centering
\includegraphics[width=1\textwidth,trim={0.25cm 0.2cm 0.05cm 0.15cm},clip]{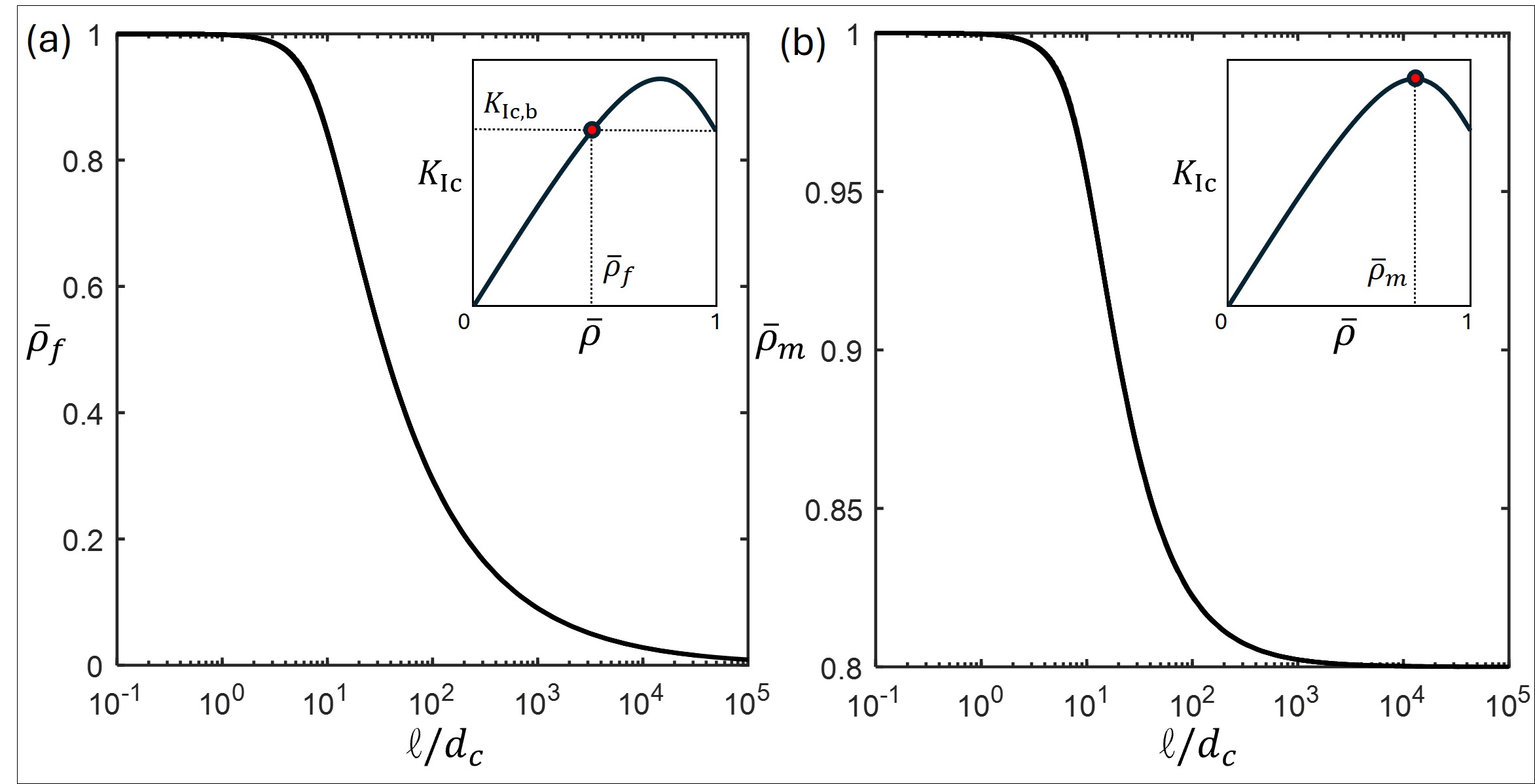}
\caption{Critical relative density results for the QB homogenized lattice model. (a) The relative density whose lattice first exceeds the base material fracture toughness is plotted against relative notch size (cell size divided by critical distance parameter, $d_c$). The inset shows this point in the $K_{\textnormal{Ic}}$ vs. $\bar\rho$ space for an example lattice with arbitrary cell size and base material properties. (b) The relative density corresponding to the lattice with the maximum fracture toughness is plotted against relative notch size. The inset shows this maximum in the $K_{\textnormal{Ic}}$ vs. $\bar\rho$ space for an example lattice with arbitrary cell size and base material properties.}\label{Fig13}
\end{figure}

The first critical relative density computed is the lowest relative density whose lattice fracture toughness is equal to the base material fracture toughness ($\bar{\rho}_f$). It is found by equating \autoref{eq:elasNotchStressQB} to the base material fracture toughness and solving for the smallest positive $\bar\rho$ which satisfies the equation. The second critical relative density is the relative density that maximizes lattice fracture toughness ($\bar{\rho}_m$). It is found by taking the first derivative of \autoref{eq:elasNotchStressQB} with respect to $\bar\rho$, setting the equation equal to zero, and solving for the critical strut thickness, $t_c$. Then, $\bar{\rho}_m$ is calculated from $\ell$ and $t_c$. Base material properties are combined as $K_{\textnormal{Ic,b}}/\sigma_{\textnormal{TS}}$ and are swept from $K_{\textnormal{Ic,b}}/\sigma_{\textnormal{TS}}=10^{-3}$ to $10^{0} \sqrt{\textnormal{m}}$ while cell size is swept from  $10^{-3}$ to $10^{2}$ mm when calculating both critical relative densities. When plotted against $\ell/d_c$, the $\bar{\rho}_f$ and $\bar{\rho}_m$ curves collapse to the two master curves shown in \autoref{Fig13}(a-b), respectively. 

When the lattice cell size is smaller than the QB fracture lengthscale ($\ell<d_c$) both $\bar{\rho}_f$ and $\bar{\rho}_m$ are approximately equal to 1 such that no lattice geometry will have a higher fracture toughness than the base material. For $\ell>d_c$, however, high-$\bar\rho$ lattices begin to outperform the base material in terms of toughness and are able to do so at lower relative densities as $\ell/d_c$ increases. For the QB PMMA lattices tested in this work, $\ell/d_c=18.1$, giving $\bar{\rho}_f=0.67$ and $\bar{\rho}_m=0.91$. The transition point at $\ell=d_c$ makes sense. The stresses ahead of the notch increase significantly when $\ell\rightarrow0$, consistent with an infinitely sharp crack tip. When $\ell > d_c$, failure is determined through circular notch fracture where the stresses decay according to \autoref{eq:elasNotchStress}. As $\ell/d_c$ continues to increase, lattice fracture toughness increases as it is dominated by $\ell$ (which controls the notch radius), while base material fracture toughness decreases as it is proportional to ${d_c}^{1/2}$. This leads to smaller and smaller values of $\bar{\rho}_f$. Interestingly, when $\ell/d_c\rightarrow\infty$, $\bar{\rho}_m\rightarrow0.8$, suggesting that the relative density which maximizes fracture toughness will never fall below 0.8. 

Designing lattice materials that meet or exceed the fracture toughness of their base material while reducing mass can enable the design of more efficient mechanical structures. These critical relative density curves can be used to choose the geometry that minimizes mass or maximizes toughness. For example, if the QB base material and cell size are fixed (fixed $\ell/d_c$), one may choose a relative density (by controlling the strut thickness) which a) minimizes mass while retaining the base material toughness ($\bar{\rho}_f$) or b) maximizes fracture toughness ($\bar{\rho}_m$). Additionally, if the amount of desired weight-saving is known (relative density is known), then a cell size, $\ell$, can be calculated to meet the fracture toughness of the base material.

\subsection{Experimental Validation}
\begin{figure}[!htb]
\centering
\includegraphics[width=1\textwidth,trim={0.3cm 0.1cm 0.1cm 0.2cm},clip]{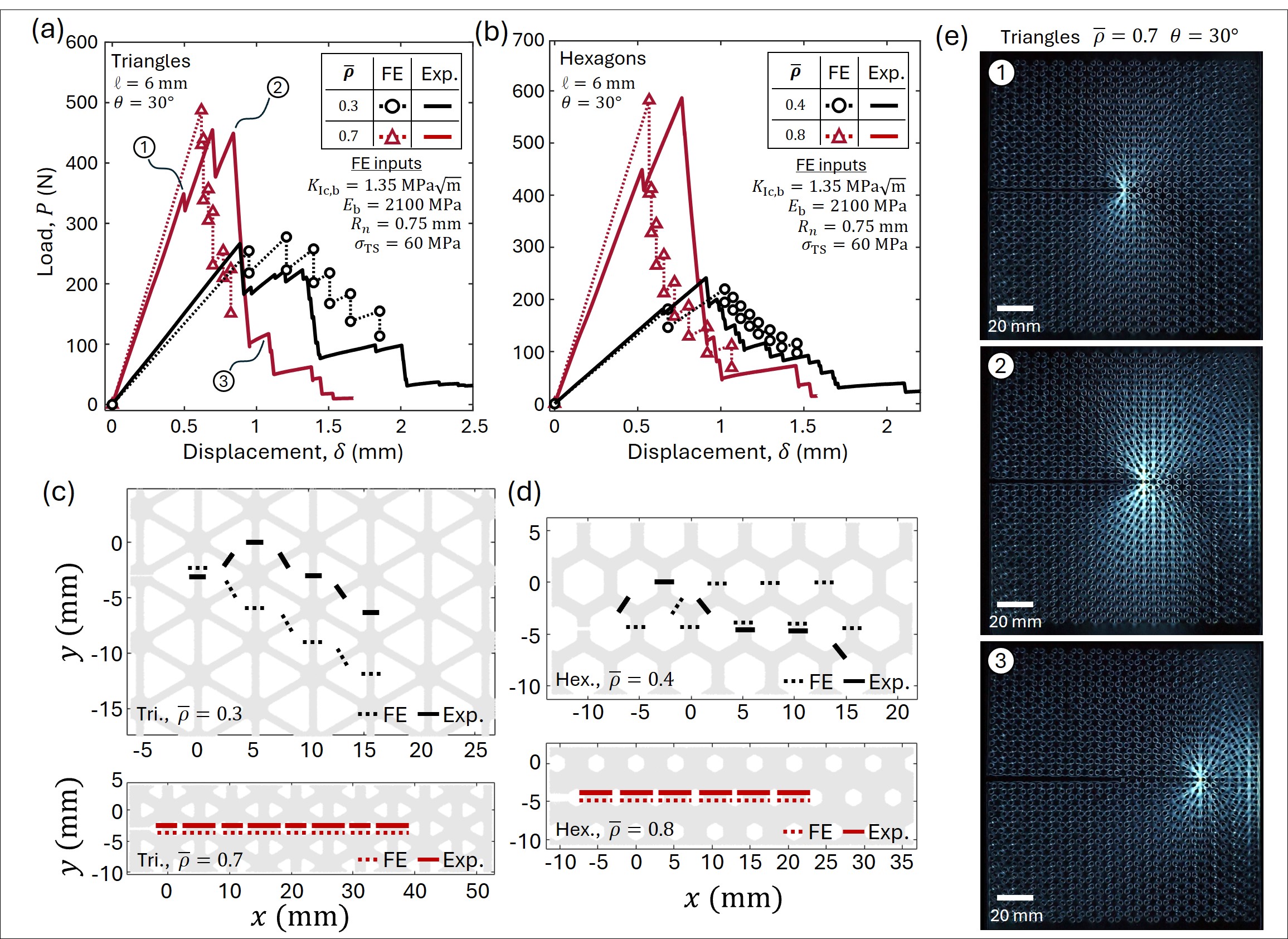}
\caption{Experimental results. Experimental and FE load-displacement curves for (a) triangular and (b) hexagonal lattices. Corresponding experimental and FE crack paths for (c) the triangular lattice tests and (d) the hexagonal lattice tests. (e) Photoelastic images of a triangular lattice with $\bar{\rho}=0.7$ and $\theta=30^\circ$ taken at three points during testing just before crack growth.}\label{Fig14}
\end{figure}

Experiments on PMMA lattices are performed following \autoref{expMethods} to validate the FE model. Results from tests on four representative lattice geometries are shown in \autoref{Fig14}, including triangular and hexagonal lattices of two relative densities and $\theta=30^\circ$. Experimental load-displacement curves are plotted alongside the FE results in \autoref{Fig14}(a-b) for triangles and hexagons, respectively. 

To slightly improve the agreement between the FE and experimental lattice results, the material properties used in the FE simulations ($E_\textnormal{b}$, $K_{\textnormal{Ic,b}}$, $\sigma_{\textnormal{TS}}$) were modified, but remain within one standard deviation of experimental tests (see \autoref{Fig14} for values used). Crucially, the real cell fillet radius, $R$, in PMMA samples was measured to be consistently 50$\%$ larger than the nominal value of 0.5 mm, likely due to the laser-cutting process, so $R=0.75$ mm for the FE comparison. 

The experimental results for all tested geometries agree well with the FE model assuming quasi-brittle failure. Initial loading stiffness, peak loads at first rupture, and the shape of the damage curve during crack propagation are all consistent with FE results across all geometries. The full experimental dataset is detailed in \ref{app4}. Slight differences in failure points along the damage curve are apparent in most samples, likely due to the stochastic nature of bulk PMMA and to small defects (e.g., geometric imperfections or surface roughness) introduced during fabrication. This leads to a variation in the strength of individual struts and alternate crack paths. The spread of damage curves among tests is shown in \autoref{FigD1} and \autoref{FigD2} for each geometry, confirming the stochastic nature of laser-cut PMMA. Still, the FE model accurately captures the fracture behavior across different cell shapes, relative densities, and orientations. Experimental and FE crack paths are shown in \autoref{Fig14}(c-d), showing good agreement in both the type of rupture (i.e., strut-based or node-based) and direction. We used photoelastic imaging to record the stress field and crack path throughout each test, as shown for one example experiment in \autoref{Fig14}(e), and in the supplementary videos for two lattices (triangular lattices with $\bar\rho=0.3,0.7$, and $\theta=30^\circ$). Brighter regions correspond to regions of high stress and concentrate around the current crack tip. For the representative high-$\bar\rho$ sample (\autoref{Fig14}(e)), we observe that a smaller stress field develops around the first crack tip at point 1 in the test than at point 2, which corresponds to a slightly lower peak load. This premature failure of the first strut is likely due to a defect in the lattice, which weakens it. However, as the test progresses to point 3, the experimental curve recovers the FE-predicted response, where the overall magnitude and shape become more consistent. A combination of strut and node-based failure is observed in the high-$\bar\rho$ triangular sample, while strut-dominated failure exists for the $\bar\rho=0.3$ sample as expected. Other experimental samples follow a similar trend, highlighting the stochastic nature of laser-cut PMMA, but remaining consistent with FE findings.

The fracture toughness from the lattice experiments are calculated in a similar manner as presented in \autoref{angComMethod} and \autoref{femodel}. The first three stable points along the damage curve of each sample are evaluated for triangular and hexagonal lattice geometries with $\theta=30^\circ$. An average fracture toughness and standard deviation are calculated for each lattice geometry and tabulated in \autoref{expKIcTable} along with the corresponding FE results with identical inputs as described in \autoref{Fig14}.

\begin{table}[!htb]
\centering
\caption{Experimental and FE fracture toughness of triangular and hexagonal lattices with $\theta=30^\circ$.}
\renewcommand{\arraystretch}{1.3}
\begin{tabular}{|c|cc|c|cc|}
\hline
\multirow{2}{*}{$\bar\rho$} & \multicolumn{2}{c|}{Triangles}                     & \multirow{2}{*}{$\bar\rho$} & \multicolumn{2}{c|}{Hexagons}                      \\ \cline{2-3} \cline{5-6} 
                            & \multicolumn{1}{c|}{$K_\textnormal{Ic,FE} \textnormal{ (MPa}\sqrt{\textnormal{m}})$}            & {$K_\textnormal{Ic,Exp} \textnormal{ (MPa}\sqrt{\textnormal{m}})$}          &                             & \multicolumn{1}{c|}{$K_\textnormal{Ic,FE} \textnormal{ (MPa}\sqrt{\textnormal{m}})$}            & {$K_\textnormal{Ic,Exp} \textnormal{ (MPa}\sqrt{\textnormal{m}})$}          \\ \hline
0.3                         & \multicolumn{1}{c|}{1.01$\pm$0.08} & 0.96$\pm$0.10 & 0.4                         & \multicolumn{1}{c|}{0.78$\pm$0.06} & 0.80$\pm$0.12 \\
0.5                         & \multicolumn{1}{c|}{1.38$\pm$0.03} & 1.61$\pm$0.14 & 0.6                         & \multicolumn{1}{c|}{1.36$\pm$0.06} & 1.54$\pm$0.12 \\
0.7                         & \multicolumn{1}{c|}{1.62$\pm$0.05} & 1.70$\pm$0.31 & 0.8                         & \multicolumn{1}{c|}{1.82$\pm$0.09} & 2.11$\pm$0.34 \\ \hline
\end{tabular}
\label{expKIcTable}
\end{table}

Fracture toughness values are consistent between FE and experimental specimens. At higher $\bar\rho$, experimental lattices have more variance across samples as stress concentrations begin to dominate failure, and defects introduced during fabrication have more influence on the fracture process. Additionally, the lowest $\bar\rho$ lattice tested has the best agreement with the FE results for both cell shapes, giving confidence in the FE model when the QB lengthscales ($r_c, d_c$) approach the strut thickness, $t$.

\section{Conclusion}
The fracture behavior of triangular and hexagonal lattices across a wide range of relative densities is investigated. Finite element simulations reveal that at higher relative densities, stress concentrations govern failure, and fracture toughness values are lower than those predicted by the classical Gibson-Ashby models as expected. Our analytical model predicts lattice fracture toughness past $\bar\rho=0.4$ remarkably well, with the triangular lattice fracture model converging to GA models at low relative densities. Stress distributions near the crack tip are significantly affected by the local geometry (which is primarily a function of relative density, orientation, and fillet radius), and crack angle, leading to a wide range of fracture properties. Lattices that achieve the highest fracture toughness at a given relative density are proficient at delocalizing these stress concentrations. Stress is shared more effectively between nodes at $\theta=30^\circ$ orientation, while increasing fillet radii delocalizes stress at each node, generating at best a 1.2$\times$ and $1.7\times$ increase in fracture toughness compared to base geometries, respectively. Additionally, adding quasi-brittle material behavior enhances the fracture toughness of the lattice by 1.6$\times$ on average as failure is delayed through localized yielding. We find from the FE results that PMMA lattices with $\ell=6$ mm cell size have higher fracture toughness than their base material starting at around $\bar\rho=0.6$ in triangular lattices and $\bar\rho=0.7$ in hexagonal lattices -- an achievement only possible when the cell size is much larger than the quasi-brittle fracture lengthscale, $d_c$. Experimental testing on laser-cut PMMA lattices confirms the trends observed in the FE studies by validating the load-displacement response measured during fracture, the macroscopic crack paths, and the measurements of fracture toughness. 

These results demonstrate the utility of lattices with high relative density, but also highlight some drawbacks. The transition to failure via stress concentration makes high-density lattices more susceptible to surface defects -- a defect present at the site of a stress concentration can cause premature failure and lead to reduced toughness. When using brittle base materials, ensuring proper surface quality during fabrication is necessary to achieve the improved toughness possible with increasing relative density. However, with appropriate geometric design and defect management, the fracture toughness of lattices can be tailored to exceed that of their base material while reducing mass. The observed trends apply to elastic-brittle and quasi-brittle lattices -- extending this framework to lattices with extensive plastic deformation remains a topic for future investigation.


\section*{Declaration of competing interests}
The authors declare that they have no known competing financial interests or personal relationships that could have appeared to influence the work reported in this paper.

\section*{Acknowledgments}
This research was funded primarily by the National Science Foundation (NSF) (award DMR-2309043) and an Ashton Fellowship from the University of Pennsylvania. 

\section*{CRediT authorship contribution statement}
\textbf{Adam Taylor:} Conceptualization, Methodology, Investigation, Formal analysis, Writing – original draft, Writing – review and editing. \textbf{Sage Fulco:} Conceptualization, Writing – review and editing, Supervision. \textbf{Kevin Turner:} Conceptualization, Methodology, Writing – review and editing, Supervision, Funding acquisition.

\newpage

\appendix

\section{Quasi-brittle Notch Stress Validation}
\label{app1}

\setcounter{figure}{0}   

\begin{figure}[!htb]
\centering
\includegraphics[width=1\textwidth,trim={0.2cm 0.2cm 0.9cm 0.2cm},clip]{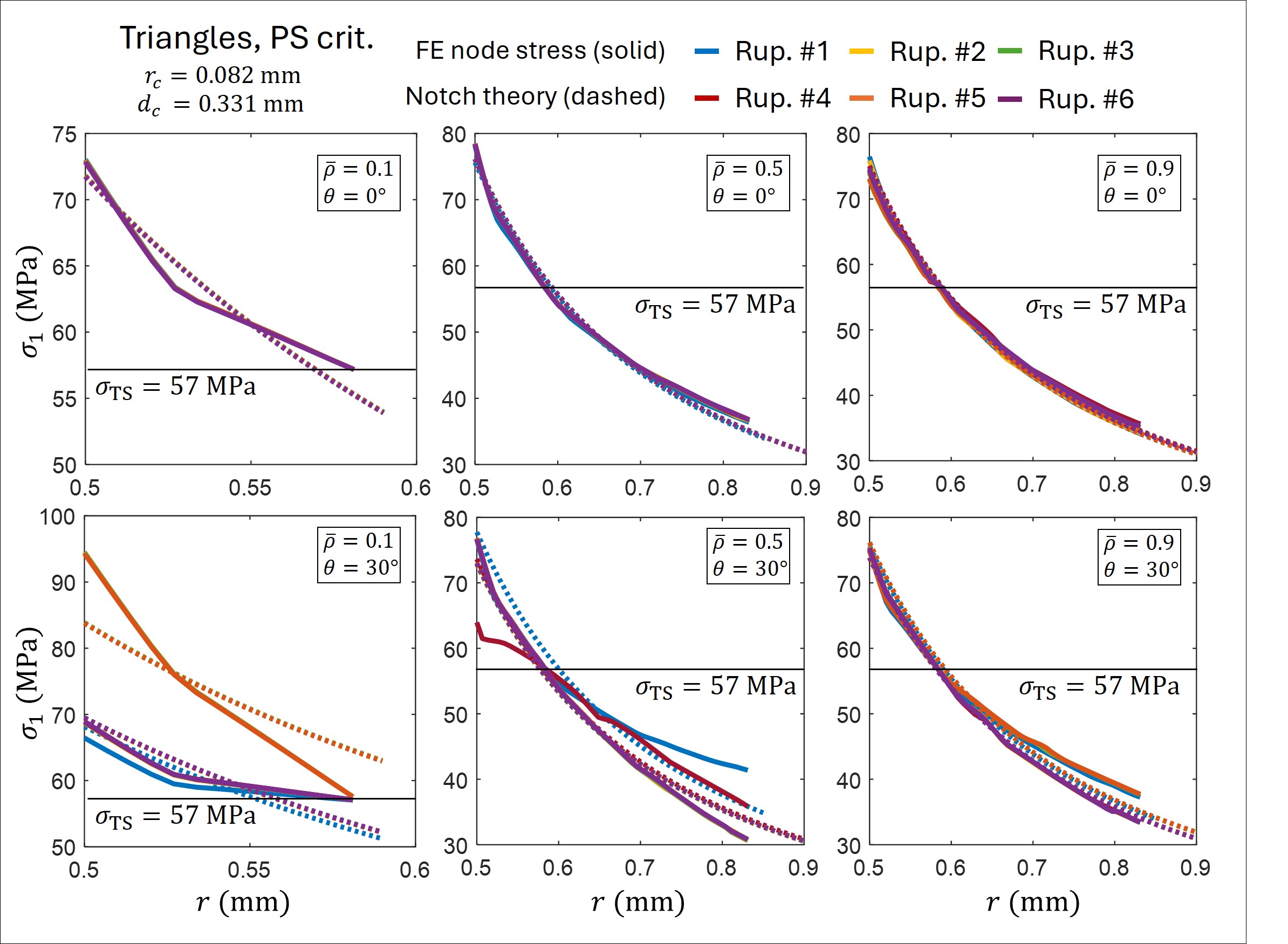}
\caption{FE elastic stress distribution at the peak stress concentration for multiple rupture events in triangular lattices. Results are scaled according to the PS criterion. Theoretical fits to the FE data are added using quasi-brittle notch fracture theory.}\label{FigA1}
\end{figure}

\begin{figure}[!htb]
\centering
\includegraphics[width=1\textwidth,trim={0.2cm 0.2cm 0.9cm 0.2cm},clip]{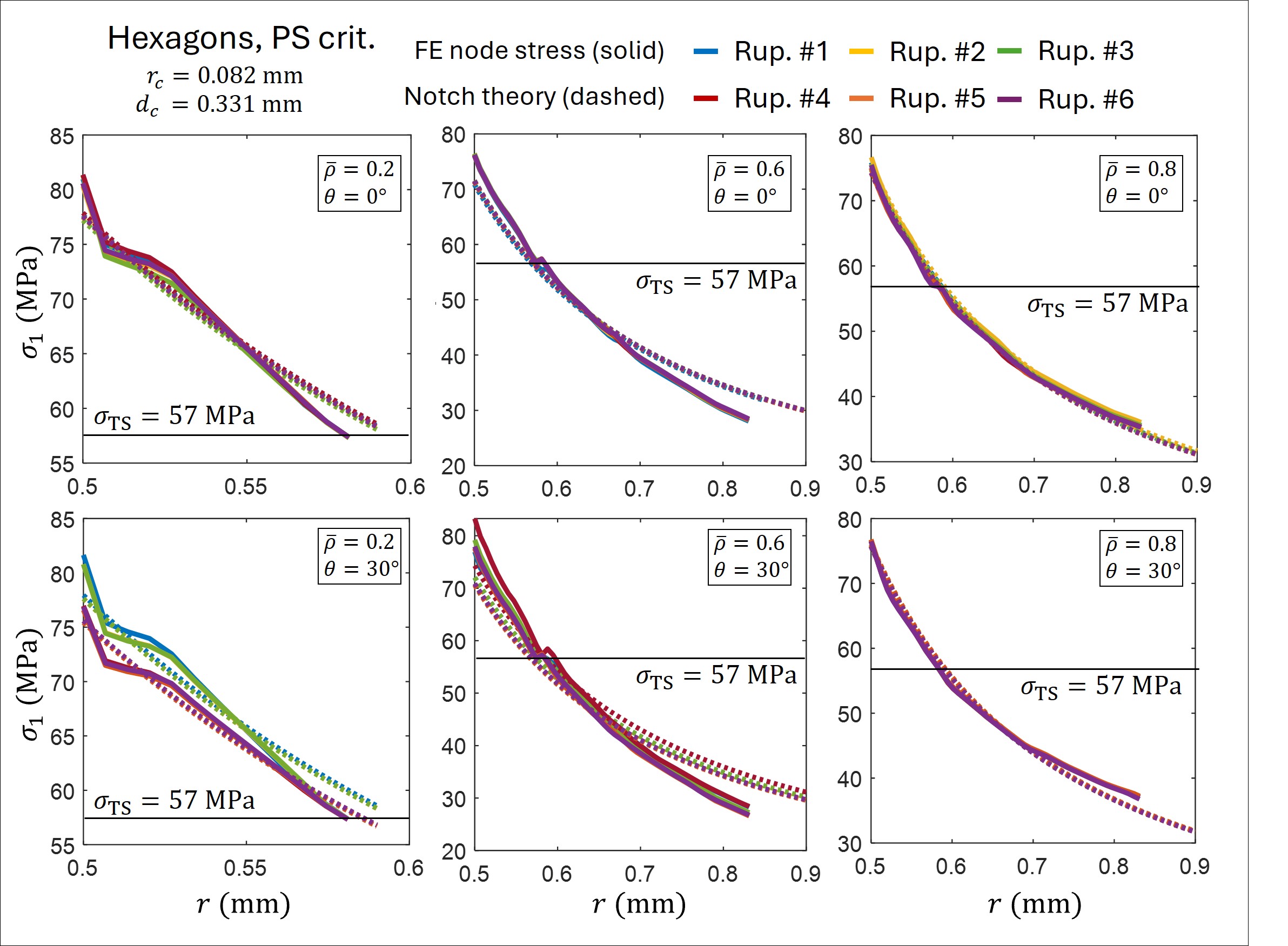}
\caption{FE elastic stress distribution at the peak stress concentration for multiple rupture events in hexagonal lattices. Results are scaled according to the PS criterion. Theoretical fits to the FE data are added using quasi-brittle notch fracture theory.}\label{FigA2}
\end{figure}

To confirm the validity of the quasi-brittle failure criterion, we compare the theoretical stress distribution ahead of a quasi-brittle notch fracture specimen to the elastic stress distribution in our FE lattice results. The FE stress distribution ahead of the dominant stress concentration for each rupture event is extracted in the direction of crack growth predicted by the MERR criterion and scaled according to the PS criterion such that the stress at the distance $r_{c}=0.082$ mm (with PMMA as the base material) from the notch edge equals the tensile strength of the material ($\sigma_\textnormal{TS}=57$ MPa). The theoretical stress distribution given by \autoref{eq:elasNotchStress} is then fit to the FE data using $K_\textnormal{I}^n$ as an effective fitting parameter and assuming $R_n$ as the fillet radius ($R=0.5$ mm). The results are compared in \autoref{FigA1} and \autoref{FigA2} for the triangular and hexagonal lattices, respectively, for three relative densities and two orientations. The FE stresses are consistent with the stresses predicted by quasi-brittle notch fracture theory for all lattice geometries, especially in geometries where the distribution is able to fully decay away from the edge. The FE toughness results using the PS criterion will be more sensitive to the exact stress distribution than those using the MS criterion, because the PS criterion evaluates a single point in the distribution rather than averaging over a length. Regardless, the FE distributions match closely to theory, giving confidence in using the quasi-brittle notch fracture theory to estimate failure in these lattices.

\section{MERR Crack Path Validation}
\label{app2}

\setcounter{figure}{0}   

\begin{figure}[!htb]
\centering
\includegraphics[width=1\textwidth,trim={0.3cm 0.2cm 0.4cm 0.2cm},clip]{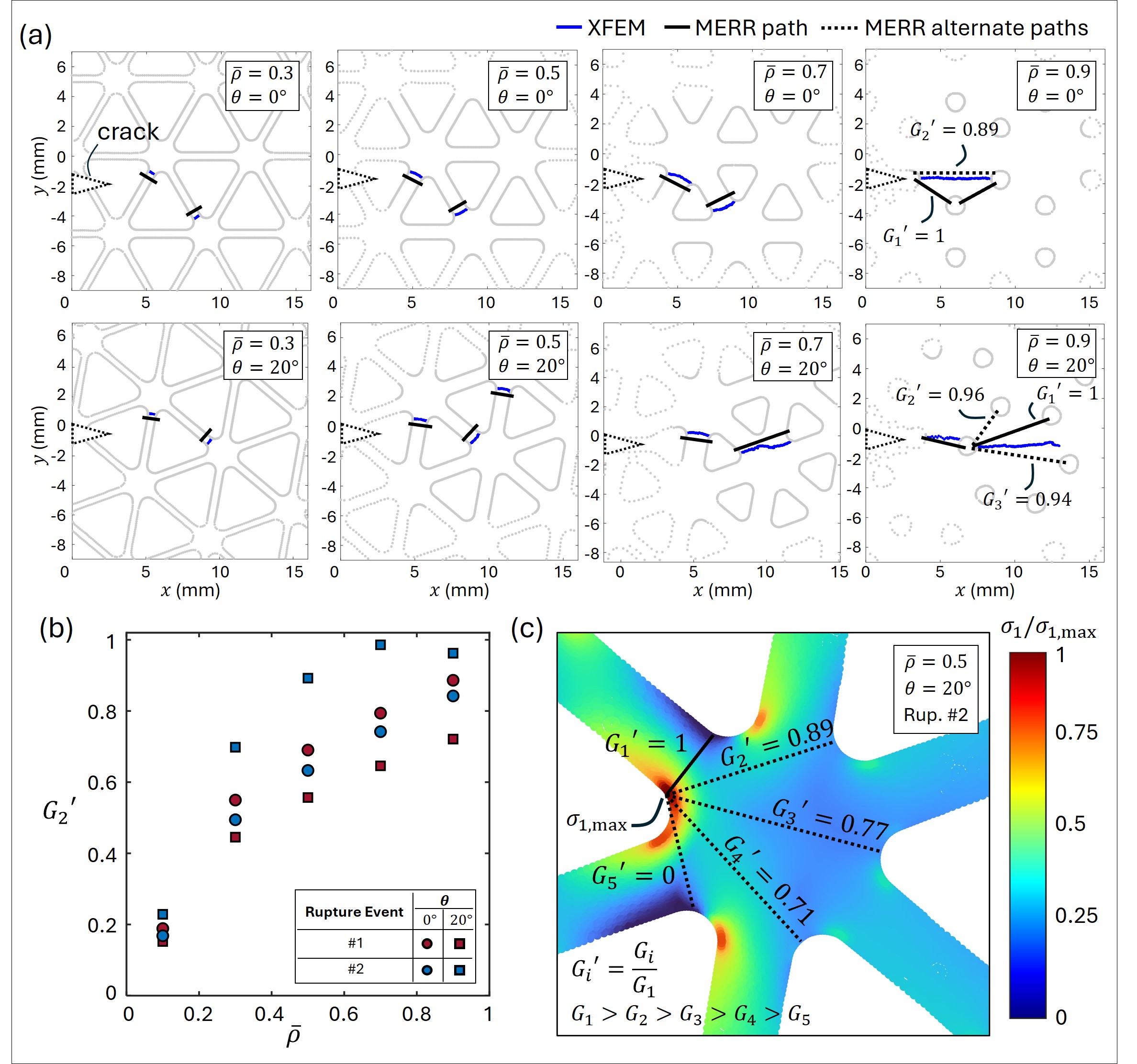}
\caption{MERR crack path validation and results. (a) Predicted crack paths according to the MERR criterion (solid black) and XFEM (blue). For cases where the two disagree, alternate MERR paths are shown with corresponding $G$ ratios. (b) The relative strain energy release rate of the second most-likely path, $G_2$', plotted against relative density for multiple rupture events and orientations. (c) All MERR crack paths and relative strain energy release rates, $G_i$', for a triangular lattice and the corresponding stress distribution in the cell.}\label{FigB1}
\end{figure}

Extended finite element method (XFEM) simulations are conducted for a subset of triangular lattice geometries to observe sharp crack growth and confirm the MERR crack path criterion. Each simulation was performed using ABAQUS' built-in XFEM crack method and defining a traction-separation law for failure. A maximum principal stress of 57 MPa and small fracture energy of $5\times10^{-4}$ $\textnormal{kJ/m}^2$ is chosen for the traction-separation law to approximate perfectly brittle fracture. A small amount of viscous dissipation is added to the model to help with convergence and stability. An element size of 0.04 mm is chosen for each model, and a vertical displacement increment of 0.004 mm is used to ensure proper formation and growth of the XFEM crack. The overall size of the lattice is reduced to 10 $\times$ 10 cells to reduce the cost of each XFEM simulation, but the study is still expected to provide insight into the validity of the MERR criterion. Four relative densities and two orientations were simulated using MERR and XFEM methods and the crack path results can be seen in \autoref{FigB1}(a). 
The MERR criterion is effective at predicting cell rupture, especially for $\bar{\rho}<0.9$, and captures both nodal and strut-based failure. At very high densities, however, the difference in strain energy release rates, $G_i$, for each potential path becomes small, leading to discrepancies with the XFEM results. The strain energy release rate of the second-most likely crack path, $G_2$, is normalized by that of the most likely path, $G_1$, to produce $G_2$' and is plotted as a function of relative density in \autoref{FigB1}(b) for two rupture events. At $\bar{\rho}=0.1$, strut failure is essentially guaranteed for any orientation, corresponding to a low $G_2$'. However, as relative density increases, the likelihood for the crack to choose a different path increases. The stresses within the cell at the crack tip for one lattice are shown in \autoref{FigB1}(c) along with all MERR alternate paths to show how the local geometry and distribution of stress affect the strain energy release rates $G_i$. The angle of each path relative to the x-axis will also affect $G$ -- the crack wants to grow along the x-axis following mode-I failure, so potential crack paths more parallel to the x-axis will generally have increased $G$. For example, the $G_2$' values for the first and second rupture in the $\theta=0^\circ$ lattices are very similar to one another for a given density as the local crack tip geometry is mirrored. However, in the $\theta=20^\circ$ lattices, the $G_2$' values differ significantly between first and second rupture as the crack tip geometry changes. The local crack tip geometry of the first rupture features strut-based failure with a crack path which is parallel to the x-axis, giving a low $G_2$'. However, it is unclear whether the second rupture event will grow through the strut or grow straight. At $\bar{\rho}=0.7$, the most energetically favorable path is through the node, and this XFEM crack matches with the MERR prediction. As density is increased to $\bar{\rho}=0.9$, all $G_2$' values are above 0.7, and the MERR criterion struggles to predict each rupture path. Despite this, the macroscopic crack path remains consistent with the XFEM result, lending confidence in the measured fracture toughnesses at high densities.

\section{Modified Compliance Method Fitting and Validation}
\label{app3}

\setcounter{figure}{0}   

\begin{figure}[!htb]
\centering
\includegraphics[width=1\textwidth,trim={0.2cm 0.2cm 0.3cm 0.1cm},clip]{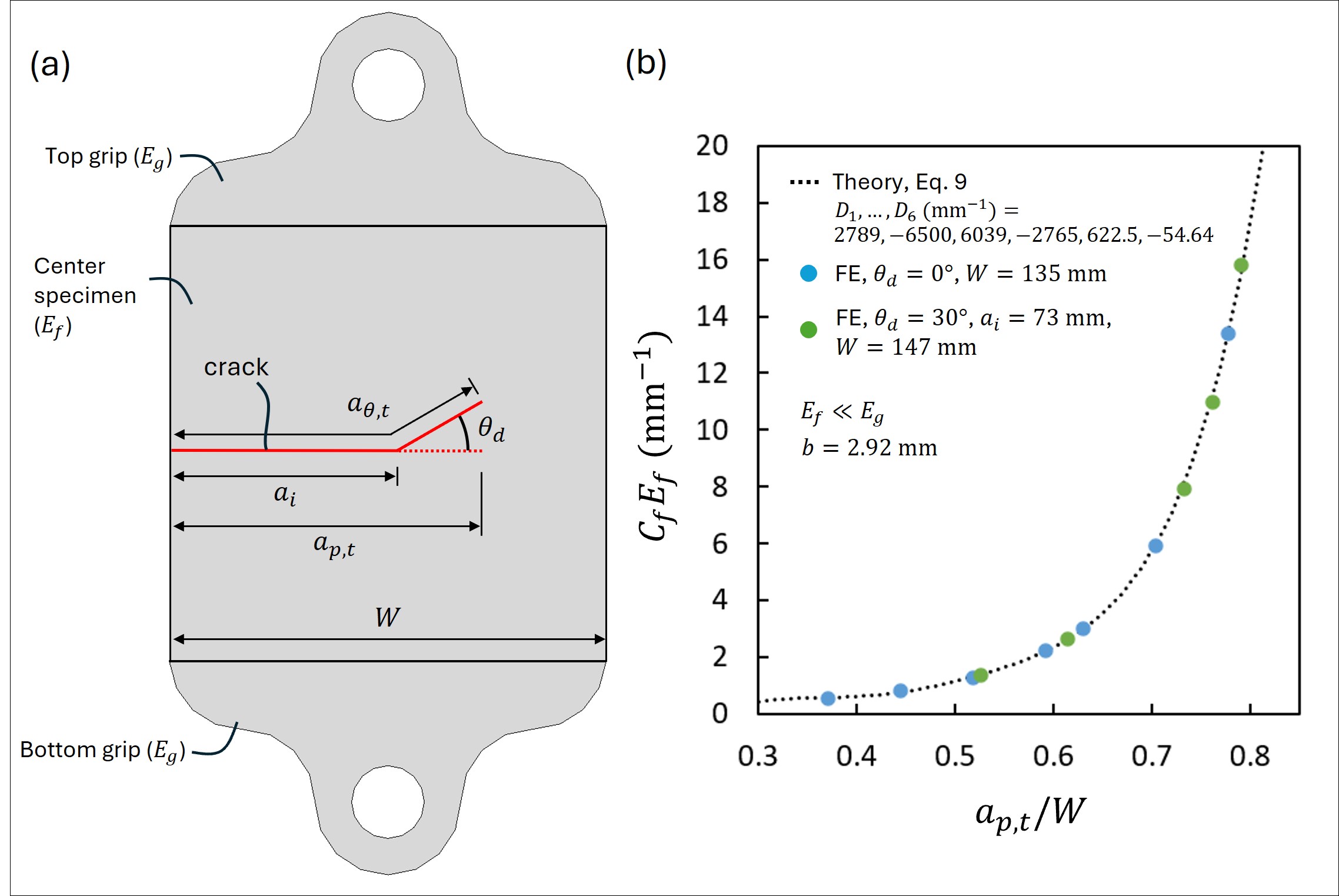}
\caption{Modified compliance method model and results. (a) Schematic of FE model used for the angled compliance method fitting showing the deflected crack geometry and overall specimen design. (b) Normalized specimen compliance, $C_fE_f$, plotted against normalized crack length, $a_{p,t}/W$, for both straight (blue) and $30^\circ$ deflected (green) cracks. Theoretical specimen compliance (\autoref{eq:CvsAcurve}) is fit to the straight crack results to determine $D_1$ through $D_6$.}\label{FigC1}
\end{figure}

A FE model of a homogeneous fracture specimen with the same overall geometry as all FE lattice specimens is constructed to extract specimen compliance as a function of crack length. The specimen geometry is shown in \autoref{FigC1}(a), including the crack geometry for a deflected crack. Separate FE models with deflected and straight cracks are tested, and their compliance is measured as a function of real crack length, assuming negligible compliance from either grip (i.e., $E_g\gg E_f$). The projected crack lengths, $a_{p,t}$, are normalized by specimen width, $W$, and plotted against measured specimen compliance times specimen effective modulus as seen in \autoref{FigC1}(b). \autoref{eq:CvsAcurve} is fit to the straight crack dataset (blue points) to extract $D_1$ through $D_6$. The deflected crack results (green points) fall along this curve as well, allowing calculation of $G_c$ for any given $a_i$ and $\theta_d$. 

\section{Additional Experimental Results}
\label{app4}

\setcounter{figure}{0}   

\begin{figure}[!htb]
\centering
\includegraphics[width=1\textwidth,trim={0.2cm 0.2cm 0.1cm 0.3cm},clip]{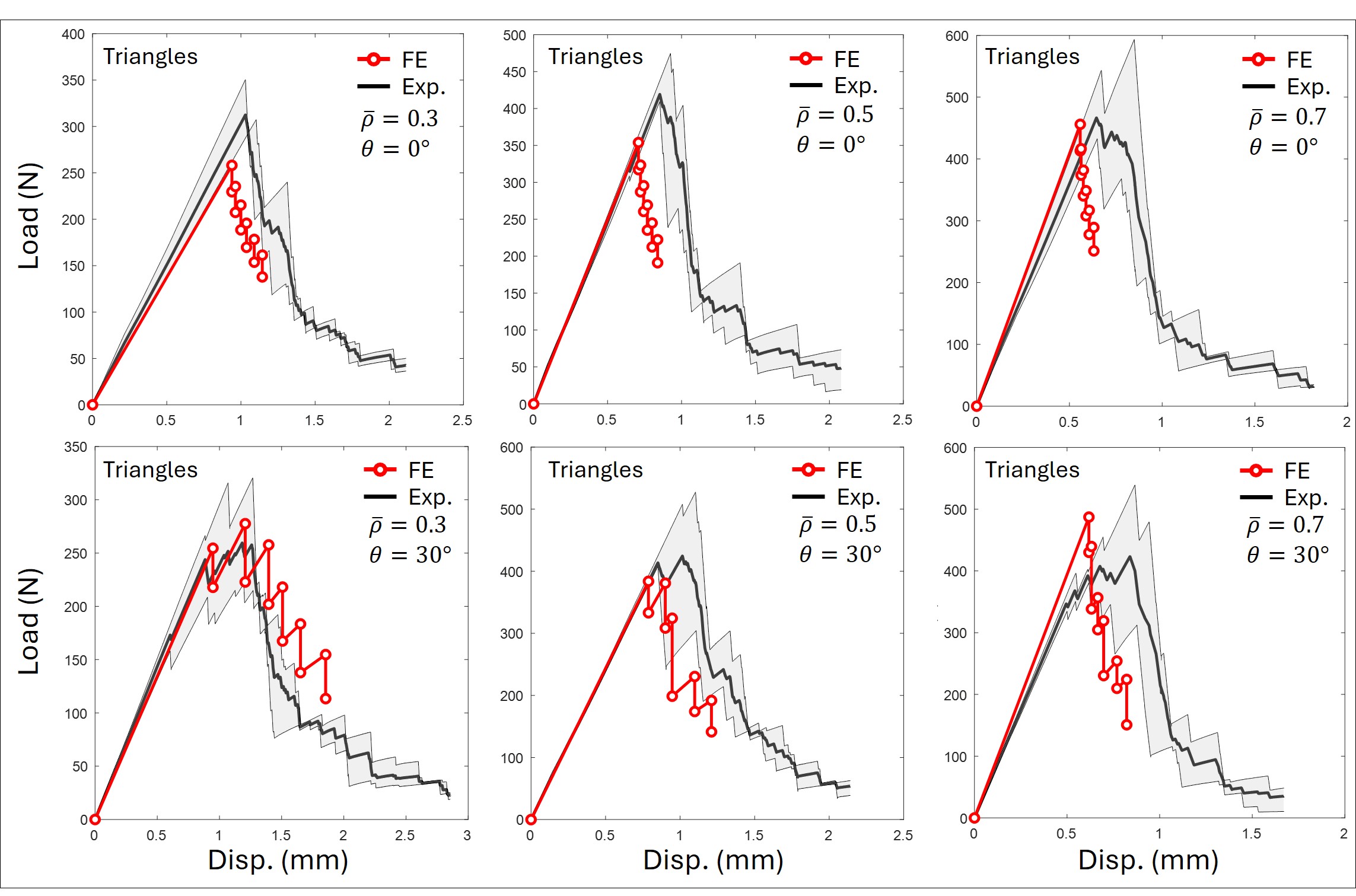}
\caption{All triangular lattice load-displacement results comparing experimental tests to the FE model assuming quasi-brittle failure. Experimental curves show the average (solid black line) and the minimum and maximum (grayed region) of three samples.}\label{FigD1}
\end{figure}

\begin{figure}[!htb]
\centering
\includegraphics[width=1\textwidth,trim={0.2cm 0.2cm 0.05cm 0.2cm},clip]{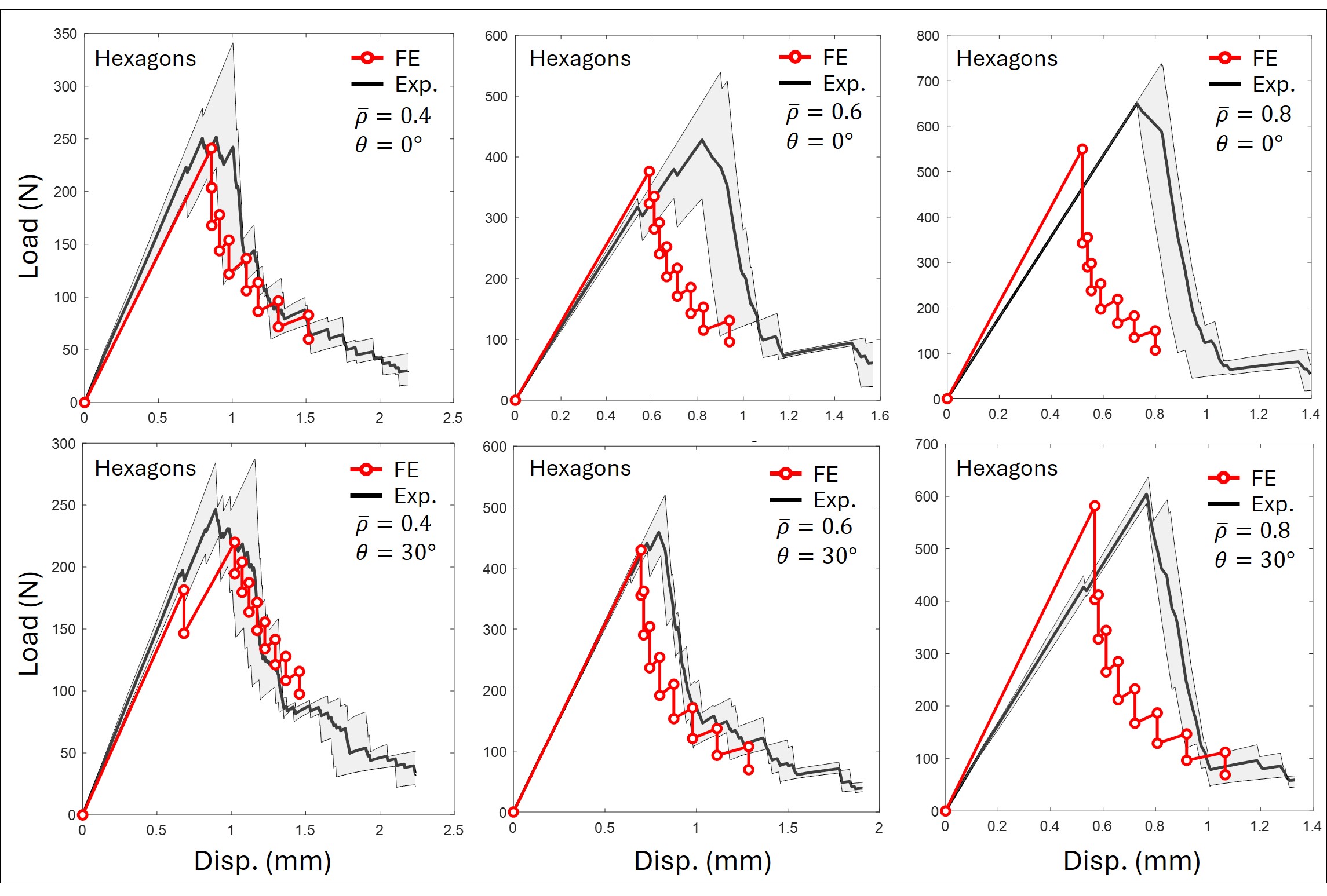}
\caption{All hexagonal lattice load-displacement results comparing experimental tests to the FE model assuming quasi-brittle failure. Experimental curves show the average (solid black line) and the minimum and maximum (grayed region) of three samples.}\label{FigD2}
\end{figure}

The full experimental load-displacement results are compared to FE for the triangular and hexagonal lattices in \autoref{FigD1} and \autoref{FigD2}, respectively. 
Once again, several FE input material properties ($E_\textnormal{b}$, $K_{\textnormal{Ic,b}}$, $\sigma_{\textnormal{TS}}$) as well as fillet radius, $R$, are modified slightly. Overall, the initial loading compliance, peak loads, and damage curves align closely with FE, giving confidence in the FE failure criteria and model. Disagreement between the two results is likely due to fabrication or material defects present within experimental samples. Clearly, different samples of the same lattice geometry can exhibit different fracture behavior, leading to variation in toughness.

\section{Quasi-brittle homogenized lattice model}
\label{app5}
The elastic-brittle homogenized lattice theory presented in \autoref{lattHomogTheory} is extended to include quasi-brittle material behavior. Instead of looking at the point stress at the notch edge as performed for elastic-brittle lattices, the elastic stress distribution is evaluated over a distance, $d_c$, from the notch edge \parencite{torabi2014closed}. Failure occurs when the average stress in this region reaches the tensile strength of the base material, $\sigma_{\textnormal{TS}}$. To find the average stress at failure, we use \autoref{eq:elasNotchStress} and integrate over the $d_c$ region as follows:
\begin{equation}
    \frac{1}{d_c}\int_{R_n}^{R_n+d_c} \sigma_{t}dr=\frac{1}{d_c}\int_{R_n}^{R_n+d_c}\frac{K_{\textnormal{I}}^{\textnormal{n}}}{2\sqrt{2\pi r}}\left[ 2+1.25\left(\frac{R_n}{r}\right)+1.5\left(\frac{R_n}{r}\right)^2+1.25\left(\frac{R_n}{r}\right)^3\right]dr
\label{eq:elasNotchStressIntQB}
\end{equation}
where $\sigma_t$ is the tangential stress parallel to loading, ${K_{\textnormal{I}}^{\textnormal{n}}}$ is the mode-I notch stress intensity factor, $R_n$ is the notch radius, and $r$ is the radius from the center of the circular notch. The left side of \autoref{eq:elasNotchStressIntQB} equals $\sigma_{\textnormal{TS}}$ at failure. Upon integrating the right side, the equation can be solved for QB lattice fracture toughness, $K_{\textnormal{Ic}}$, producing: 
\begin{equation}
    K_{\textnormal{Ic}}=K_{\textnormal{I}}^{\textnormal{n}}H=2\sqrt{2\pi}d_c\sigma_{\textnormal{TS}}H\left[4\sqrt{R_n+d_c}-\frac{2.5R_n}{\sqrt{R_n+d_c}}-\frac{{R_n}^2}{(R_n+d_c)\sqrt{R_n+d_c}}-\frac{0.5{R_n}^3}{(R_n+d_c)^2\sqrt{R_n+d_c}}\right]^{-1}
\label{eq:elasNotchStressQB}
\end{equation}
where $H$ is the parameter which accounts for the homogenization applied to the lattice. For a triangular lattice, $H$ is equal to $\bar{\rho}$ following the equivalent stress analysis presented in \autoref{lattHomogTheory}. The notch radius for the triangular lattice is the same as in the elastic-brittle derivation (\autoref{eq:effRadTri}) while $d_c=0.331$ mm for PMMA calculated from \autoref{eq:dc}. Quasi-brittle triangular lattice fracture toughness is plotted in \autoref{Fig12} as a function of $\bar\rho$ and shows good agreement with the FE results. 

\section*{Data Availability}
Data will be made available on request.




\printbibliography





\end{document}

\endinput